%% file: main_s216.tex
\documentclass[ twocolumn,floatfix,superscriptaddress,showpacs,nofootinbib,preprintnumbers,secnumarabic,amssymb, nobibnotes, aps, prd]{revtex4-2}
\usepackage[utf8]{inputenc}
\usepackage{graphicx}
\usepackage{latexsym,amsmath,amssymb,amsthm,lmodern,float,url}
\usepackage{tikz}
\usepackage{quantikz}
\usepackage{natbib}
\usepackage{color}
\usepackage{microtype}
\usepackage{import}
\usepackage{bbold}
\usepackage[plain]{fancyref}
\usepackage{varioref}
\usepackage{slashed}
\usepackage{bbm}
\usepackage{multirow}
\usepackage{tikz}
\usepackage{scrextend}
\usepackage{braket}
\usetikzlibrary{shapes}
\usetikzlibrary{positioning}
\usepackage[normalem]{ulem}
\usepackage{adjustbox}
\usepackage{balance}
\usepackage[dvipsnames]{xcolor}

\usepackage[colorlinks=true,backref=false, linktocpage=true,
citecolor=RoyalBlue,urlcolor=RoyalBlue,linkcolor=RoyalBlue,pdfpagemode=UseOutlines]{hyperref}
\hypersetup{%
  bookmarksnumbered=true,
  pdftitle = {},
  pdfsubject = {},
  pdfauthor = {},
  pdfkeywords = {}
}

\usepackage{orcidlink}

\DeclareMathOperator{\tr}{Tr}
\DeclareMathOperator{\re}{Re}

\newcommand{\bi}{\mathbb{BI}}
\newcommand{\bo}{\mathbb{BO}}
\newcommand{\btt}{\mathbb{BT}}
\DeclareMathOperator{\dsd}{\Sigma(36\times3)}
\DeclareMathOperator{\dsa}{\Sigma(72\times3)}
\DeclareMathOperator{\dsb}{\Sigma(216\times3)}
\DeclareMathOperator{\dsc}{\Sigma(360\times3)}

\definecolor{blugrn}{RGB}{0,158,115}

\newcommand{\sqms}{\affiliation{Superconducting and Quantum Materials System Center (SQMS), Batavia, IL, 60510, USA.}}
\newcommand{\fnal}{\affiliation{Fermi National Accelerator Laboratory, 
Batavia, IL 60510, USA}}

\begin{document}
\preprint{FERMILAB-PUB-25-0674-SQMS-STUDENT-T}

\title{Primitive Quantum Gates for an \texorpdfstring{$SU(3)$}{SU(3)} Discrete Subgroup: \texorpdfstring{$\Sigma(72\times3)$}{S(216)}}

\author{Sebastian Osorio Perez\,\orcidlink{0009-0002-3247-4975}}
\email{sebasop@umd.edu}
\fnal
\affiliation{Department of Physics, University of Maryland, College Park, MD 20742, USA}

\author{Edison M. Murairi\,\orcidlink{0000-0002-1639-6308}}
\email{emurairi@fnal.gov}
\fnal
\sqms

\author{Erik J. Gustafson\,\orcidlink{0000-0001-7217-5692}}
\email{egustafson@usra.edu}
\affiliation{Universities Space Research Association, Research Institute for Advanced Computer Science (RIACS) at NASA Ames Research Center, Moffett Field, California, 94035, USA}

\author{Henry Lamm\,\orcidlink{0000-0003-3033-0791}}
\email{hlamm@fnal.gov}
\fnal
\sqms

\begin{abstract}
We construct a primitive gate set for the digital quantum simulation of a discrete subgroup of $SU(3)$: the 216-element $\Sigma(72\times3)$. The necessary primitives are the inversion gate, the group multiplication gate, the trace gate, and the group Fourier transform, for which we provide qubit decompositions.  The resulting fault-tolerant T gate costs for a fiducial calculation of shear viscosity would require about $10^{12}$ T gates which compares favorably to other modern estimates.
\end{abstract}

\maketitle
\section{Introduction}
While classical computers have demonstrated great success in computing properties of gauge theories with Monte Carlo methods, this approach fails for real time dynamics of the system or at finite fermion density due to the sign problems~\cite{Aarts_2016,Gattringer:2016kco,Alexandru:2020wrj,Troyer:2004ge}. Quantum computers can study these problems but the limited quantum resources require efficient implementations of the infinite-dimensional bosonic degrees of freedom and their interactions.

To this end, several bosonic digitizations have been proposed in the past decade. Some quantize in the representation (electric field) basis, and impose a cut off on the maximal representation~\cite{Unmuth-Yockey:2018ugm,Unmuth-Yockey:2018xak, Klco:2019evd, Farrell:2023fgd, Farrell:2024fit, Ciavarella:2024lsp,Illa:2024kmf,Ciavarella:2021nmj, Bazavov:2015kka, Zhang:2018ufj,PhysRevD.99.114507,Bauer:2021gek,Grabowska:2022uos,Buser:2020uzs,Bhattacharya:2020gpm,Kavaki:2024ijd,Calajo:2024qrc,Murairi:2022zdg,Zohar:2015hwa,Zohar:2012xf,Zohar:2012ay,Zohar:2013zla} while the $q-$deformed formulation~\cite{Zache:2023dko,Zache:2023cfj} replaces the gauge group by a quantum group. Other digitizations reformulate the degrees of freedom, for example: loop-string-hadron formulation~\cite{Davoudi:2024wyv,Raychowdhury:2018osk,Kadam:2023gli,Davoudi:2020yln,Mathew:2022nep}, light-front quantization~\cite{Kreshchuk:2020dla,Kreshchuk:2020aiq,Kreshchuk:2020kcz}, conformal truncation~\cite{Liu:2020eoa}, strong-coupling and large-$N_c$ expansions~\cite{Fromm:2023bit,Ciavarella:2024fzw}, fuzzy gauge theories from non-commutative geometry~\cite{Alexandru:2023qzd}, and quantum link models~\cite{Brower:1997ha,Singh:2019jog,Singh:2019uwd,Wiese:2014rla,Brower:1997ha,Brower:2020huh,Mathis:2020fuo,Halimeh:2020xfd, budde2024quantum, osborne2024quantum, Osborne:2023rzx,Luo:2019vmi}. Across this breadth, the typical estimate for qubits per local degree of freedom is $\mathcal{O}(10)$ for reasonable approximations of $SU(3)$, but establishing their connections to the continuum quantum field theory is non-trivial~\cite{Alexandru:2023qzd,ORLAND1990647,Singh:2019jog}.  Most approaches considered only qubit methods, but there has been an increasing interest in qudits~\cite{Gustafson:2021jtq,Gustafson:2021qbt,Popov:2023xft,kurkcuoglu2022quantum,Calajo:2024qrc,Gonzalez-Cuadra:2022hxt, Illa:2024kmf, Zache:2023cfj,Ale:2024uxf,Pardo:2024edu}. 

In this work, we continue the study of the discrete subgroup approximation~\cite{Bender:2018rdp,Hackett:2018cel,Alexandru:2019nsa,Yamamoto:2020eqi,Ji:2020kjk,Haase:2020kaj,Carena:2021ltu,Armon:2021uqr,Gonzalez-Cuadra:2022hxt,Charles:2023zbl,Irmejs:2022gwv,Gustafson:2020yfe, Bender:2018rdp,Hackett:2018cel,Alexandru:2019nsa,Yamamoto:2020eqi,Ji:2020kjk,Haase:2020kaj,Carena:2021ltu,Armon:2021uqr,Gonzalez-Cuadra:2022hxt,Charles:2023zbl,Irmejs:2022gwv, Hartung:2022hoz,Carena:2024dzu,Zohar:2014qma,Zohar:2016iic, Davoudi:2024wyv, Mueller:2024mmk} which has its roots in the classical Euclidean lattice field theory~\cite{Creutz:1979zg,Creutz:1982dn,Bhanot:1981xp,Petcher:1980cq,Bhanot:1981pj,Hackett:2018cel,Alexandru:2019nsa,Ji:2020kjk,Ji:2022qvr,Alexandru:2021jpm,Carena:2022hpz,Weingarten:1980hx,Weingarten:1981jy}. The discrete subgroup approximation has several advantages. First, it provides a mapping to integers while preserving a remnant of gauge symmetry; replacing fixed- or floating-point arithmetic with modular arithmetic. The discrete symmetry also allows for coupling the gauge redundancy to quantum error correction schemes~\cite{rajput2021quantum,Carena:2024dzu,Spagnoli:2024mib,Kurkcuoglu:2025gik,Haruna:2025piy}. Third, the quantum memory requirements are fixed for a given discrete subgroup and improvement in truncation error require only including additional interactions and thus circuit depth~\cite{Carena:2022kpg,Alexandru:2021jpm}.  This contrasts with other digitizations were reducing error requires more memory and deeper circuits.   Finally, discrete subgroup approximations can be studied classically~\cite{Alexandru:2021jpm,Assi:2024pdn}; something other digitizations can struggle with due to sign problems. 

On the other hand, the discrete subgroup approximation requires trade offs.  For nonabelian theories, one is limited to the small number of \emph{crystal-like} subgroups with the correct center.  For $SU(3)$ there are only four choices: $\Sigma(\phi\times3)$ for $\phi=36,72,216,360$. On the lattice, there are first-order phase transitions at some \emph{freezeout} coupling $\beta_f$ (or lattice spacing $a_f$) that separates the discrete group from the continuum limit of the continuous group. These $\beta_f$, alongside other important properties of the smallest crystal-like subgroups of $SU(3)$ discrete groups are found in Tab.~\ref{tab:groupcomp}. We agree with previous work on these groups~\cite{Bhanot:1981xp}, but report higher precision values. Theoretical work has shown that in the continuum the field theory of discrete groups is related to the symmetry-broken phase of continuous groups coupled to Higgs bosons in particular representations~\cite{Kogut:1980qb,romers2007discrete,Fradkin:1978dv,Harlow:2018tng,Horn:1979fy}. Recent work has begun to quantify this connection for non-abelian groups~\cite{Assi:2024pdn}. 

\begin{table}
\caption{\label{tab:groupcomp} Parameters of a crystal-like subgroups of $SU(3)$. $\Delta S$ is the gap between $\mathbb{1}$ and the nearest neighbors to it. $\mathcal{N}$ is number of group elements that neighbor the $\mathbb{1}$.}
\begin{center}
\begin{tabular}
{c| c c c c c}
\hline\hline
$G$ & $\Delta S$ & $\mathcal{N}$ & $\beta^{2+1d}_f$ & $\beta^{3+1d}_f$ \\
\hline
$\dsd$&$\frac{2}{3}$&18&3.78(2)&2.52(3)\\
$\dsa$&$\frac{2}{3}$&54&4.223(1)&3.18(3)\\
\hline\hline
\end{tabular}
\end{center}
\end{table}

\begin{figure}[!th]
\centering
    \includegraphics[width=0.9\linewidth]{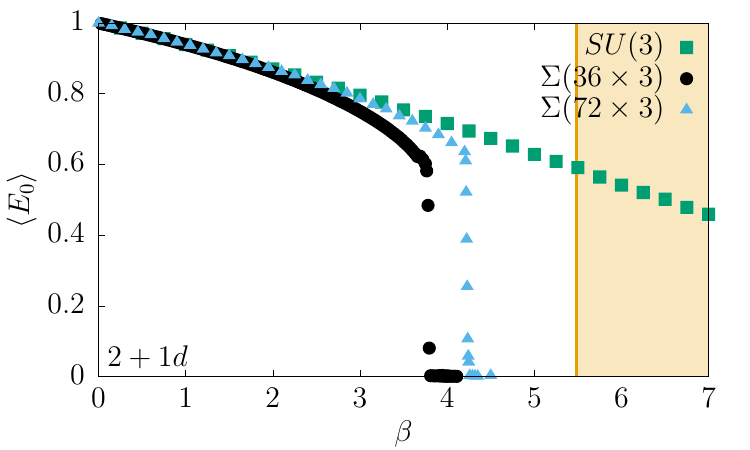}
    \includegraphics[width=0.9\linewidth]{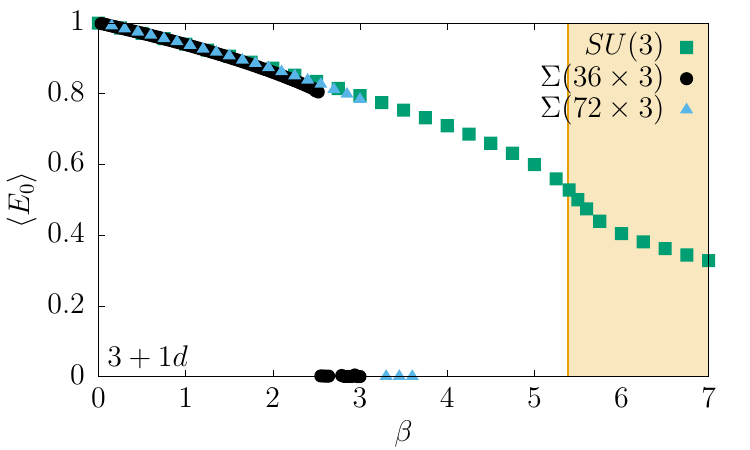}
    \caption{\label{fig:freezing}Euclidean calculations of the expectation value of the plaquette $\langle E_0\rangle$ as a function of Wilson coupling $\beta$  on $8^d$ lattices for (top) $(2+1)$-dimensions (bottom) $(3+1)$-dimensions. The shaded region indicates the scaling regime $\beta\geq \beta_s$.} 
    \label{fig:lattice-energy}
\end{figure}

In this work, we consider the crystal-like subgroup of a $SU(3)$ -- $\dsa$ which has 216 elements. Dependence of the lattice energy on coupling in the Wilson action are shown in Fig.~\ref{fig:lattice-energy}.  The elements of $\dsa$ can be encoded into a register consisting of 9 qubits. This work constitutes another rung in the ladder of discrete subgroups for quantum simulation following previous work on smaller groups.  Subgroups of $SU(2)$ studied are: the $2N$-element dihedral groups $D_N$~\cite{Bender:2018rdp,Lamm:2019bik,Alam:2021uuq,Fromm:2022vaj}, the 8-element $\mathbb{Q}_8$~\cite{Gonzalez-Cuadra:2022hxt}, the 24-element $\btt$~\cite{Gustafson:2022xdt}, the 48 element $\bo$~\cite{Gustafson:2023kvd}, the 120 element $\bi$~\cite{Lamm:2024jnl}.  For $SU(3)$, primitive gates have only been derived for the 108-element $\dsd$~\cite{Gustafson:2024kym}.  

From Fig.~\ref{fig:lattice-energy}, we see that freeze-out occurs before the scaling regime.  This implies that the Kogut-Susskind Hamiltonian would be insufficient for $\dsa$ to approximate the continuum limit of $SU(3)$ and including additional interactions is necessary~\cite{Alexandru:2019nsa,Ji:2020kjk,Ji:2022qvr,Alexandru:2021jpm,Assi:2024pdn}.  Together with arguments relating lattice actions to lattice Hamiltonians~\cite{Carena:2021ltu,Clemente:2022cka,Funcke:2022opx,Carena:2022hpz,Crippa:2024cqr}, this suggests that discrete subgroups would be adequate for quantum simulations. 

This paper is organized as follows. In Sec.~\ref{sec:group}, we present the group-theoretic details of $\Sigma(36 \times 3)$ and the digitization scheme. A summary of the qudit gates used is found in Sec~\ref{sec:gates}. Sec.~\ref{sec:primitive-gates} briefly reviews the necessary primitive gates for gauge theories, and then explicit circuits for $\dsa$ are derived in the following sections: the inversion gate in Sec.~\ref{sec:inverse}, the multiplication gate in Sec.~\ref{sec:multiplication}, the trace gate in Sec.~\ref{sec:trace}, and the Fourier transform gate in Sec.~\ref{sec:fourier}. Using these gates, we discuss the resource costs for simulating pure gauge $SU(3)$ in $3+1d$ in Sec.~\ref{sec:resources}.  We conclude and discuss future work in Sec.~\ref{sec:conc}.

\section{Group Properties}
\label{sec:group}
\begin{table}\setlength{\tabcolsep}{1.1pt}
\caption{Character table of $\dsa$ with $\omega=e^{2\pi i/3}$ adapted from~\cite{Grimus:2010ak}.
\label{tab:cts216}}
\begin{tabular}{c|ccc|c|ccc|ccc|ccc|ccc}
Size & 1&1&1 & 24 & 9&9&9 & 18 & 18 &18 &18 &18 &18 &18 &18 &18\\
Ord & 1 & 3 & 3 & 3 & 2 & 6 & 6 & 4 & 12 & 12 & 4 & 12 & 12 &
12 & 12 & 4 \\
\hline \hline
$\mathbf{1}^{(0)}$ & $1$ & $1$ & $1$ & $1$ & $1$ 
& $1$ & $1$ & $1$
& $1$ & $1$ & $1$
& $1$ & $1$ & $1$ & $1$ & $1$ \\
$\mathbf{1}^{(1)}$ & $1$ & $1$ & $1$ & $1$ & $1$ 
& $1$ & $1$ & $\text{-}1$
& $\text{-}1$ & $\text{-}1$ & $1$ 
& $1$ & $1$ & $\text{-}1$ & $\text{-}1$ & $\text{-}1$ \\
$\mathbf{1}^{(2)}$ &
$1$ & $1$ & $1$ & 
$1$ &
$1$ & $1$ & $1$ &
$1$ & $1$ & $1$ &
$\text{-}1$ & $\text{-}1$ & $\text{-}1$ &
$\text{-}1$ & $\text{-}1$ & $\text{-}1$ \\
$\mathbf{1}^{(3)}$ &
$1$ & $1$ & $1$ & 
$1$ &
$1$ & $1$ & $1$ &
$\text{-}1$ & $\text{-}1$ & $\text{-}1$ &
$\text{-}1$ & $\text{-}1$ & $\text{-}1$ &
$1$ & $1$ & $1$ \\
\hline
$\mathbf{2}$ &
$2$ & $2$ & $2$ & 
$2$ &
$\text{-}2$ & $\text{-}2$ & $\text{-}2$ &
$0$ & $0$ & $0$ &
$0$ & $0$ & $0$ &
$0$ & $0$ & $0$ \\
\hline
$\mathbf{3}^{(0)}$ &
$3$ & $3\omega$ & $3\omega^2$ & 
$0$ &
$\text{-}1$ & $\text{-}\omega$ & $\text{-}\omega^2$ &
$1$ & $\omega$ & $\omega^2$ &
$1$ & $\omega$ & $\omega^2$ &
$\omega$ & $\omega^2$ & $1$ \\
$\mathbf{3}^{(1)}$ &
$3$ & $3\omega$ & $3\omega^2$ & 
$0$ &
$\text{-}1$ & $\text{-}\omega$ & $\text{-}\omega^2$ &
$\text{-}1$ & $\text{-}\omega$ & $\text{-}\omega^2$ &
$1$ & $\omega$ & $\omega^2$ &
$\text{-}\omega$ & $\text{-}\omega^2$ & $\text{-}1$ \\
$\mathbf{3}^{(2)}$ &
$3$ & $3\omega$ & $3\omega^2$ & 
$0$ &
$\text{-}1$ & $\text{-}\omega$ & $\text{-}\omega^2$ &
$1$ & $\omega$ & $\omega^2$ &
$\text{-}1$ & $\text{-}\omega$ & $\text{-}\omega^2$ &
$\text{-}\omega$ & $\text{-}\omega^2$ & $\text{-}1$ \\
$\mathbf{3}^{(3)}$ &
$3$ & $3\omega$ & $3\omega^2$ & 
$0$ &
$\text{-}1$ & $\text{-}\omega$ & $\text{-}\omega^2$ &
$\text{-}1$ & $\text{-}\omega$ & $\text{-}\omega^2$ &
$\text{-}1$ & $\text{-}\omega$ & $\text{-}\omega^2$ &
$\omega$ & $\omega^2$ & $1$ \\
$\mathbf{3}^{(0)*}$ &
$3$ & $3\omega^2$ & $3\omega$ & 
$0$ &
$\text{-}1$ & $\text{-}\omega^2$ & $\text{-}\omega$ &
$1$ & $\omega^2$ & $\omega$ &
$1$ & $\omega^2$ & $\omega$ &
$\omega^2$ & $\omega$ & $1$ \\
$\mathbf{3}^{(1)*}$ &
$3$ & $3\omega^2$ & $3\omega$ & 
$0$ &
$\text{-}1$ & $\text{-}\omega^2$ & $\text{-}\omega$ &
$\text{-}1$ & $\text{-}\omega^2$ & $\text{-}\omega$ &
$1$ & $\omega^2$ & $\omega$ &
$\text{-}\omega^2$ & $\text{-}\omega$ & $\text{-}1$ \\
$\mathbf{3}^{(2)*}$ &
$3$ & $3\omega^2$ & $3\omega$ & 
$0$ &
$\text{-}1$ & $\text{-}\omega^2$ & $\text{-}\omega$ &
$1$ & $\omega^2$ & $\omega$ &
$\text{-}1$ & $\text{-}\omega^2$ & $\text{-}\omega$ &
$\text{-}\omega^2$ & $\text{-}\omega$ & $\text{-}1$ \\
$\mathbf{3}^{(3)*}$ &
$3$ & $3\omega^2$ & $3\omega$ & 
$0$ &
$\text{-}1$ & $\text{-}\omega^2$ & $\text{-}\omega$ &
$\text{-}1$ & $\text{-}\omega^2$ & $\text{-}\omega$ &
$\text{-}1$ & $\text{-}\omega^2$ & $\text{-}\omega$ &
$\omega^2$ & $\omega$ & $1$ \\
\hline
$\mathbf{6}$ &
$6$ & $6\omega$ & $6\omega^2$ & 
$0$ &
$2$ & $2\omega$ & $2\omega^2$ &
$0$ & $0$ & $0$ &
$0$ & $0$ & $0$ &
$0$ & $0$ & $0$ \\
$\mathbf{6^{\ast}}$ &
$6$ & $6\omega^2$ & $6\omega$ & 
$0$ &
$2$ & $2\omega^2$ & $2\omega$ &
$0$ & $0$ & $0$ &
$0$ & $0$ & $0$ &
$0$ & $0$ & $0$ \\
\hline
$\mathbf{8}$ &
$8$ & $8$ & $8$ & 
$\text{-}1$ &
$0$ & $0$ & $0$ &
$0$ & $0$ & $0$ &
$0$ & $0$ & $0$ &
$0$ & $0$ & $0$ \\
\end{tabular}
\end{table}
In this section, we overview the properties of $\dsa$. 
This discussion includes the generating representation used, the irreducible representations, and how the digitization is mapped onto qubit-based platforms. The group has the following generators:

\begin{align}
    \label{eq:matrices}
    C = \begin{pmatrix} 1 & 0 & 0\\0 & \omega & 0\\0 & 0 & \omega^2\end{pmatrix};\qquad&
    V = \frac{1}{\sqrt{3}i}\begin{pmatrix}
        1 & 1 & 1\\1 & \omega & \omega^2\\1 & \omega^2 & \omega\end{pmatrix}\notag\\ E =\begin{pmatrix} 0 & 1 & 0\\0 & 0 & 1\\
    1 & 0 & 0\end{pmatrix};\qquad&X = \frac{1}{\sqrt{3}i}\begin{pmatrix} 1 & 1 & \omega^2\\1 & \omega & \omega\\\omega&1&\omega
        \end{pmatrix}
\end{align}
with $\omega=e^{2\pi i / 3}$.
These matrices are in the faithful SU(3) representations of $\dsa$ and have the properties
\begin{align}
    \label{eq:dsarules}
    X C =  C^2 E X, & &  X E = \omega^2 C E X \notag\\
    X V = \omega^2 E V^3 X && 
    C^3 = E^3 = V^4 = X^4 = \mathbbm{1}.
\end{align}
With these, any element $g$ of $\dsa$ can be written as
    \begin{align}
    \label{eq:s216orderedproduct}
        g = \omega^p C^q E^r V^{2s + t} X^u,
    \end{align}
    where $0\leq p, q, r \leq 2$ and $0\leq s, t, u\leq 1$.  For hybrid qudit platforms, the natural encoding to minimize errors would be three qutrits and three qubits~\cite{Gustafson:2023swx}. For qubit platforms, we expand each of $p,q,r,y$ into two binary variables. Thus, Eq.~(\ref{eq:s216orderedproduct}) becomes
\begin{equation*}
g = \omega^{p_1+2p_2}C^{q_1+2q_2}E^{r_1+2r_2}V^{2s + t}X^u,
\end{equation*}
where each variable is binary. It is necessary to restrict the variables due to excess of qubit states: $\lambda_1 + \lambda_2 \leq 1$ for $\lambda=p,q,r$. Thus, a $\dsa$-register requires $9$ qubits and is given by the binary encoding in the $\ket{u t s r_1 r_0 q_1 q_0 p_1 p_0}$. We use $\ket{N}$ to denote this encoding where $N$ is the integer representation of the binary string. For example,
\begin{align}
\frac{i}{\sqrt{3}}\begin{pmatrix}
    \omega^2&\omega&\omega\\
    1&\omega&1\\
    1&1&\omega
\end{pmatrix}&=\omega^{1+2\times0}C^{0+2\times1}E^{1+2\times0}V^{2\times0+1}X^1\notag\\&\rightarrow\ket{110011001}=\ket{409}
\end{align}

The group $\dsa$ has 16 irreducible representations (irreps): $\mathbf{1}^{(a)}$, $\mathbf{2}$, $\mathbf{3}^{(a)}$, $\mathbf{3}^{(a),*}$, $\mathbf{6}$, $\mathbf{6}^*$, and $\mathbf{8}$ where $0\leq a \leq 3$. The bold faced numbers indicate the dimension and the superscripts differentiate the various irreps in a given dimension.
Using character table (Tab.~\ref{tab:cts216}) and Eq.~(\ref{eq:dsarules}) one can identify a presentation denoted $\rho_{i}(g)$ for each irrep $i$. These are for the 1d irreps:
\begin{align}
    \rho_{\mathbf{1}^{(0)}}(g) = 1; && \rho_{\mathbf{1}^{(1)}}(g) = (-1)^t;\notag\\
    \rho_{\mathbf{1}^{(2)}}(g) = (-1)^{u};&&\rho_{\mathbf{1}^{(3)}}(g) = (-1)^{t + u};
\end{align}
the 2d irrep:
\begin{align}
    \label{eq:dsairreps}
    \rho_{\mathbf{2}}(g) = \begin{pmatrix}
        0 & -i\\-i& 0
    \end{pmatrix}^{2s + t} \begin{pmatrix}
        0 & 1\\-1& 0
    \end{pmatrix}^u
    \end{align}
Using $\omega$, $C$, $E$, $V$, and $X$ from Eq. (\ref{eq:matrices}), the 3d irreps are    
    \begin{align}
    \rho_{\mathbf{3}^{(0)}}(g) &= \omega^p C^q E^r V^{2s + t} X^u \notag\\
    \rho_{\mathbf{3}^{(1)}}(g) &= \omega^p C^q E^r (-V)^{2s + t} X^u \notag\\
    \rho_{\mathbf{3}^{(2)}}(g) &= \omega^p C^q E^r V^{2s + t} (-X)^u \notag\\
    \rho_{\mathbf{3}^{(3)}}(g) &= \omega^p C^q E^r (-V)^{2s + t} (-X)^u.\end{align}
    The 3d conjugate representations correspond to $V,X\rightarrow \bar{V},\bar{X}$.
The 6d irreducible representation is: \begin{align}
    &\rho_{\mathbf{6}}(\omega) = e^{2\pi i / 3};\notag\\&\rho_{\mathbf{6}}(C) = \rho_{\mathbf{3}^{(0)}}(C) \oplus (\omega^2\rho_{\mathbf{3}^{(0)}}(C^2));\notag\\&\rho_{\mathbf{6}}(E) = \rho_{\mathbf{3}^{(0)}}(E)\oplus\rho_{\mathbf{3}^{(0)}}(E^2);\notag\end{align}
    \begin{align}
    &\rho_{\mathbf{6}}(V) = {\setlength{\arraycolsep}{1pt}\scriptstyle -\frac{1}{3}\begin{pmatrix} 1 & 1 & 1 & \sqrt{2} & \sqrt{2} & \sqrt{2} \\ 1 & \omega^2 & \omega & \sqrt{2} \omega & \sqrt{2} \omega^2 & \sqrt{2} \\ 1 & \omega & \omega^2 & \sqrt{2} \omega^2 & \sqrt{2} \omega & \sqrt{2} \\ \sqrt{2} & \sqrt{2} \omega & \sqrt{2} \omega^2 & \omega+1 & \omega^2+1 & -1 \\ \sqrt{2} & \sqrt{2} \omega^2 & \sqrt{2} \omega & \omega^2+1 & \omega+1 & -1 \\ \sqrt{2} & \sqrt{2} & \sqrt{2} & -1 & -1 & -1 \\ \end{pmatrix}};\notag\end{align}
    
    \begin{align}&\rho_{\mathbf{6}}(X) = {\setlength{\arraycolsep}{1pt}\scriptstyle -\frac{1}{3}\begin{pmatrix} 1 & 1 & \omega & \sqrt{2} & \sqrt{2} \omega^2 & \sqrt{2} \omega^2 \\ 1 & \omega^2 & \omega^2 & \sqrt{2} \omega & \sqrt{2} \omega & \sqrt{2} \omega^2 \\ \omega^2 & 1 & \omega^2 & \sqrt{2} \omega & \sqrt{2} \omega^2 & \sqrt{2} \omega \\ \sqrt{2} & \sqrt{2} \omega & \sqrt{2} & \omega+1 & -1 & -\omega^2 \\ \sqrt{2} \omega & \sqrt{2} & \sqrt{2} & \omega+1 & \omega+1 & -1 \\ \sqrt{2} \omega & \sqrt{2} \omega & \sqrt{2} \omega^2 & -\omega & -1 & -1 \\ \end{pmatrix}};\notag
\end{align}
for its conjugate $\rho_{\mathbf{6}^*}(g) = (\rho_{\mathbf{6}}(g))^*$. The 8d irrep is 
\begin{widetext}
\begin{gather}
    {\setlength{\arraycolsep}{1pt}\scriptstyle\rho_{\mathbf{8}}(\omega) = 1;\;~\rho_{\mathbf{8}}(C) = \begin{pmatrix}
        -\frac{1}{2} & -\frac{\sqrt{3}}{2} & 0\\
        \frac{\sqrt{3}}{2} & -\frac{1}{2} & 0\\
        0 & 0 & 1
    \end{pmatrix}\oplus\begin{pmatrix}
        -\frac{1}{2} & \frac{\sqrt{3}}{2} \\-\frac{\sqrt{3}}{2} & -\frac{1}{2}
    \end{pmatrix}\oplus \begin{pmatrix}
        -\frac{1}{2} & -\frac{\sqrt{3}}{2} & 0\\
        \frac{\sqrt{3}}{2} & -\frac{1}{2} & 0\\
        0 & 0 & 1
    \end{pmatrix};\;\rho_{\mathbf{8}}(E) = \begin{pmatrix}
        0 & 0 & 0 & 0 & 1 & 0 & 0 & 0\\
        0 & 0 & 0 & 0 & 0 & 1 & 0 & 0\\
        0 & 0 & -\frac{1}{2} & 0 & 0 & 0 & 0 & \frac{\sqrt{3}}{2}\\
        0 & 0 & 0 & 0 & 0 & 0 & 1 & 0\\
        1 & 0 & 0 & 0 & 0 & 0 & 0 & 0\\
        0 & -1 & 0 & 0 & 0 & 0 & 0 & 0\\
        0 & 0 & 0 & 1 & 0 & 0 & 0 & 0\\
        0 & 0 & -\frac{\sqrt{3}}{2} & 0 & 0 & 0 & 0 & -\frac{1}{2}
    \end{pmatrix}};\notag\\{\setlength{\arraycolsep}{0.7pt} \scriptstyle~\rho_{\mathbf{8}}(V) = \begin{pmatrix} \frac{1}{6} & -\frac{1}{2 \sqrt{3}} & \frac{1}{2} & \frac{1}{6} & \frac{1}{2 \sqrt{3}} & -\frac{1}{3} & \frac{1}{\sqrt{3}} & \frac{1}{2 \sqrt{3}} \\ \frac{1}{2 \sqrt{3}} & -\frac{1}{2} & -\frac{1}{2 \sqrt{3}} & -\frac{1}{2 \sqrt{3}} & -\frac{1}{2} & 0 & 0 & \frac{1}{2} \\ \frac{1}{2} & \frac{1}{2 \sqrt{3}} & 0 & \frac{1}{2} & -\frac{1}{2 \sqrt{3}} & \frac{1}{2} & \frac{1}{2 \sqrt{3}} & 0 \\ \frac{1}{6} & \frac{1}{2 \sqrt{3}} & \frac{1}{2} & \frac{1}{6} & -\frac{1}{2 \sqrt{3}} & -\frac{1}{3} & -\frac{1}{\sqrt{3}} & \frac{1}{2 \sqrt{3}} \\ -\frac{1}{2 \sqrt{3}} & -\frac{1}{2} & \frac{1}{2 \sqrt{3}} & \frac{1}{2 \sqrt{3}} & -\frac{1}{2} & 0 & 0 & -\frac{1}{2} \\ -\frac{1}{3} & 0 & \frac{1}{2} & -\frac{1}{3} & 0 & \frac{2}{3} & 0 & \frac{1}{2 \sqrt{3}} \\ -\frac{1}{\sqrt{3}} & 0 & -\frac{1}{2 \sqrt{3}} & \frac{1}{\sqrt{3}} & 0 & 0 & 0 & \frac{1}{2} \\ \frac{1}{2 \sqrt{3}} & -\frac{1}{2} & 0 & \frac{1}{2 \sqrt{3}} & \frac{1}{2} & \frac{1}{2 \sqrt{3}} & -\frac{1}{2} & 0 \\ \end{pmatrix};\;
 \rho_{\mathbf{8}}(X) = \begin{pmatrix} \frac{1}{6} & -\frac{1}{2 \sqrt{3}} & \frac{1}{2} & -\frac{1}{3} & 0 & -\frac{1}{3} & -\frac{1}{\sqrt{3}} & \frac{1}{2 \sqrt{3}} \\ \frac{1}{2 \sqrt{3}} & -\frac{1}{2} & -\frac{1}{2 \sqrt{3}} & \frac{1}{\sqrt{3}} & 0 & 0 & 0 & \frac{1}{2} \\ \frac{1}{2} & \frac{1}{2 \sqrt{3}} & 0 & 0 & \frac{1}{\sqrt{3}} & -\frac{1}{2} & \frac{1}{2 \sqrt{3}} & 0 \\ \frac{1}{6} & \frac{1}{2 \sqrt{3}} & -\frac{1}{2} & -\frac{1}{3} & -\frac{1}{\sqrt{3}} & -\frac{1}{3} & 0 & \frac{1}{2 \sqrt{3}} \\ \frac{1}{2 \sqrt{3}} & \frac{1}{2} & \frac{1}{2 \sqrt{3}} & 0 & 0 & \frac{1}{\sqrt{3}} & 0 & \frac{1}{2} \\ \frac{2}{3} & 0 & 0 & \frac{1}{6} & -\frac{1}{2 \sqrt{3}} & \frac{1}{6} & -\frac{1}{2 \sqrt{3}} & -\frac{1}{\sqrt{3}} \\ 0 & 0 & \frac{1}{\sqrt{3}} & \frac{1}{2 \sqrt{3}} & -\frac{1}{2} & -\frac{1}{2 \sqrt{3}} & \frac{1}{2} & 0 \\ \frac{1}{2 \sqrt{3}} & -\frac{1}{2} & 0 & -\frac{1}{\sqrt{3}} & 0 & \frac{1}{2 \sqrt{3}} & \frac{1}{2} & 0 \\ \end{pmatrix}.}\notag
\end{gather}
\end{widetext}

$\mathbf{1}^{(a)}$ and $\mathbf{2}$ are connected with the finite subgroup $\mathbb{Q}_8$ of $SU(2)$. Further, the various 3d irreps are connected to the $\mathbb{Q}_8$ structure. $\mathbf{6}$ mixes a subset of the $\Sigma(36\times3)$ 3d irreps while $\mathbf{8}$  mixes the two 4d irreps of $\Sigma(36\times3)$~\cite{Assi:2024pdn}. 

\section{Qubit \& Qutrit Gates}
\label{sec:gates}

The implementation of our group primitive gates requires defining a set of quantum gate operations. 
    We choose a universal set which includes some redundancy in order to compactly represent our circuits. We use the Pauli gates ($p=X$, $Y$, $Z$) and their arbitrary rotations generalizations $R_p(\theta)=e^{ip\theta / 2}$. How these are decomposed in terms of the $T=\text{diag}(1, e^{i\pi / 4})$ gate becomes relevant to fault-tolerant resource estimations. We further use a few two-qubit gates. The SWAP gate exchanges the state of two qubits, while the CNOT gate is defined by
        \begin{equation*}
            \text{CNOT} \ket{a}\otimes\ket{b} = \ket{a}\otimes\ket{b\oplus a},
        \end{equation*}
where $\oplus$ indicates addition modulo 2. We further use the multi-controlled gates such as the C$^n$NOT and CSWAP (Fredkin) gate. The effect of C$^2$NOT (Toffoli) on a quantum state is
        \begin{equation*}
            \text{C}^2\text{NOT} \ket{a}\otimes\ket{b}\otimes\ket{c}=\ket{a}\otimes\ket{b}\otimes\ket{c \oplus ab}.
        \end{equation*}
The CSWAP gate exchanges two qubits based on the control, and written in terms of modular arithmetic is
        \begin{equation*}
        \begin{split}
            \text{CSWAP}\ket{a}\otimes\ket{b}\otimes\ket{c}=&\ket{a}\otimes\ket{b(1\oplus a)\oplus ac}\\&\otimes\ket{c(1\oplus a)\oplus a b}.\\
            \end{split}
        \end{equation*}
Since several of the generators have order 3, it is useful to consider hybrid encodings with qutrits as an intermediate step. As such, we use a number of qutrit gates in this paper. The single-qudit rotations  are denoted by $R_{\alpha}^{(i, j)}(\theta)$, where $\alpha=\lbrace X,~Y,~Z\rbrace$ and indicate a Pauli rotation between levels $i$ and $j$. 
Further, we use the two-qudit entangling gate,
\begin{equation}
\label{eq:qutritcnot}
    C_a^i X_{b}^{(j,k)}~,
\end{equation}
which corresponds to the CNOT operation controlled on state $i$ of qutrit $a$, and targets qutrit $b$ with an $X$ operation between the levels $j$ and $k$. Another important qudit gate, $\chi$ represents the increment gate, which takes
\begin{equation}
    \chi\ket{a}=\ket{a\oplus_d 1}
\end{equation}
where $d$ indicates the dimension of the qudit as well as the modulus of the addition.  In this work, we only consider the qutrit $\chi$:
\begin{equation}
    \chi=\begin{pmatrix}
        0&0&1\\
        1&0&0\\
        0&1&0\\
    \end{pmatrix}
\end{equation}
which can be decomposed into two-qubit registers as
\begin{equation}
    \chi\ket{a}\ket{b}=(C_bX_a)(C_aX_b)(X\otimes\mathbb{1})\ket{a}\ket{b}
\end{equation}
We also for conciseness consider the $\textsc{CSum}$ gate:
\begin{equation}
    \textsc{CSum}_{a, b}|i\rangle_a|j\rangle_b = |i\rangle_a |i \oplus_3 j\rangle_b.
\end{equation}
with a control qudit $a$ and target qutrit $b$ and can be constructed from $C_{a}^{i}X_{b}^{(j,k)}$ by~\cite{Di:2011cvl}:
 \begin{align}
     \textsc{CSum}_{a, b} &= (C_a^1 X_b^{(0,1)} C_a^1 X_b^{(1,2)})(C_a^2 X_b^{(1,2)} C_a^2 X_b^{(0,1)}).
 \end{align}
 Further generalizations used are the multi-controlled versions of these gates e.g.
$C_a^iC_b^jX_c^{(k,l)}$ and
\begin{equation}
    \textsc{CCSum}_{a,b,c}|i\rangle_a|j\rangle_b|k\rangle_c = |i\rangle_a|j\rangle_b|k\oplus_3 ij\rangle_c.
\end{equation}
With these, there are several decompositions from qutrit-qubits operators to pure qubit ones summarize in App. \ref{app:toffolistaircase}. Specifically, the number of control qutrits can be reduced with the Toffoli staircase decomposition~\cite{2019arXiv190401671B,Chuang:1996hw,PhysRevA.52.3457}.

\section{Overview of Primitive Gates}
\label{sec:primitive-gates}

\begin{figure*}[!t]
\centering
\includegraphics[width=0.9\linewidth]{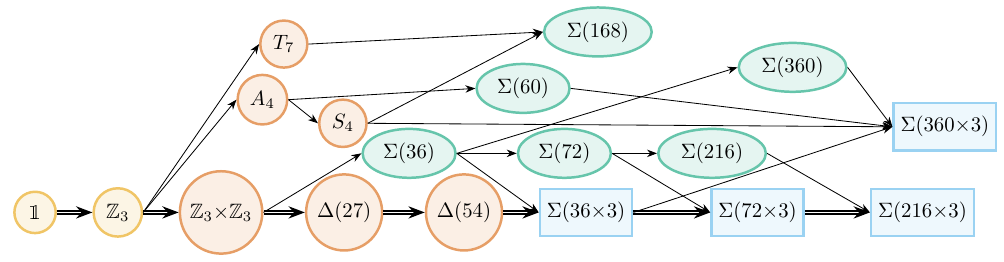}
\caption{Lattice of subgroups of $SU(3)$.  Each path represents a possible decomposition of $\mathfrak{U}_{FFT}^{G}$.  The double-line path indicates the one explored in this work.}
\label{fig:su3path}
\end{figure*}

In Ref.~\cite{Lamm:2019bik}, the Hamiltonian simulations of lattice gauge theories can be constructed from a set of group-theoretic primitive gates. This reduces the problem of simulation to constructing these primitive gates and estimating their cost. One choice of primitive gates is: inversion ($\mathfrak{U}_{-1}$), multiplication ($\mathfrak{U}_{\times}$), trace ($\mathfrak{U}_{\tr}$), Fourier transformation ($\mathfrak{U}_{F}$), Fourier phase ($\mathfrak{U}_{\phi}(\theta)$), and gauge-matter interaction ($\mathfrak{U}_{gm}(\theta)$). In this work we will consider all but the gauge-matter interaction, which we leave for future work. 

$\mathfrak{U}_{-1}$ and $\mathfrak{U}_{\times}$ permute group element basis states by
\begin{align}
    \label{eq:inversion}
    \mathfrak{U}_{-1}|g\rangle = |g^{-1}\rangle
\end{align}
and 
\begin{align}
    \label{eq:multiplication}
    \mathfrak{U}_{\times}|g\rangle\otimes|h\rangle = |g\rangle|gh\rangle.
\end{align}
This left multiplication is sufficient however an explicit right multiplication gate can further reduce costs~\cite{Carena:2022kpg}.
 
Hamiltonian simulations of lattice gauge theories also require applying phases proportional to the real part of traces:
    \begin{align}
        \label{eq:traceop}
        \mathfrak{U}_{\tr}(\theta)|g\rangle = e^{i\theta\rm{Re(Tr(}g))}.
    \end{align}

The final two gates of this set are $\mathfrak{U}_{F}$ and $\mathfrak{U}_{\phi}(\theta)$. For a finite group $G$, the \emph{discrete Fourier transform} (DFT) can be defined as $\hat f$, of a function $f$ over a finite $G$ is
\begin{eqnarray}
\hat{f}(\rho) = \sum_{g \in G}\sqrt{\frac{d_{\rho}}{|G|}}  f(g) \rho(g),
\label{eqn:Fourier-group}
\end{eqnarray}
where $\vert G \vert$ is the size of the group, $d_{\rho}$ is the dimensionality of the irrep $\rho$. With the dual $\hat{G}$ is the set of irreps of $G$, $\mathfrak U_{F}$ acts on a single $G$-register with some amplitudes $f(g)$ to rotate it into the irrep basis:
\begin{equation}
\label{eq:uft}
\mathfrak U_{FT} \sum_{g \in G} f(g)\left|g \right>
=
\sum_{\rho \in \hat G} \hat f(\rho)_{ij} \left|\rho,i,j\right>.
\end{equation}

Naively constructing the DFT requires $O(|G|)$ gates. Instead we construct a fast Fourier transform (FFT), reducing the complexity to  $O(|G| \log(G))$.  Such quantum FFTs $\mathfrak{U}^{G}_{FFT}$ have been constructed for a number of finite groups~\cite{kitaev,NC,hales2000improved,Hoyer:1997qc,beals1997quantum,Pueschel:1998zzo,Murairi:2024xpc}. The procedure for deriving $\mathfrak{U}^{G}_{FFT}$ involves extending Fourier transforms along a series of nested subgroups~\cite{Pueschel:1998zzo,Hoyer:1997qc, moore2003generic}. There are numerous paths that can be taken, given by the lattice of subgroups (See Fig~\ref{fig:su3path}).

Depending on the specifics of the discrete group, different constructions are possible, but here we follow P\"uschel et al.'s method~\cite{Pueschel:1998zzo}.
The algorithm works by appending a list of elements, denoted a \textit{right transversal}, to an existing group whose Fourier transform is known, e.g. $\mathbb{Z}_3$, and then using the effects of the right transversal to extend the irreps of the subgroup to those of a parent group, e.g. $\mathbb{Z}_3\times\mathbb{Z}_3$ or $A_4$.

To do this, one decomposes $\mathfrak{U}_{FFT}^G$ into six subroutines at each recursive step.  They are: $\mathfrak{U}_{FFT}^H$ of a subgroup $H\in G$, two irrep permutations ($P$ and $C$), a twiddle matrix ($\mathfrak{D}$), a cyclic FFT over the transversal ($DFT$), and a phase-kickback operation ($\Phi^H$).

When performing $\mathfrak{U}_{FFT}^H$, the irrep basis is often not aligned with that of $G$, thus one uses $P$ and $C$ to permute the subgroup irreps. $P$ permutes the irreps such that they are easily identifiable.  Then $C$ permutes irreps which are not invariant under conjugation by the transversal, i.e. $\rho^{a}(t g t^{-1}) \neq \rho^{a}$.  This reduces the cost and complexity for the later phase-kickback operation, $\Phi$. 

\begin{figure}[ht]
\includegraphics[width=\linewidth]{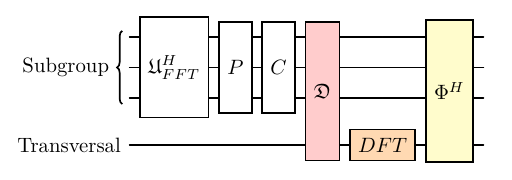}
\caption{Schematic of the $\mathfrak{U}_{FFT}^{G}$ provided in Ref. \cite{Pueschel:1998zzo}.}
\label{fig:genericfouriertransform}
\end{figure}

The twiddle operation, $T$ is controlled by a transversal register and performs two functions. It either mixes conjugate irreps to make a larger dimensional irrep or transforms an existing irrep into the desired representation. This is followed by $DFT$ on the transversal register which mixes the irreps and ensures that the transversal component is moved to the irrep basis. The final step involves a phase kickback that applies certain phases to irreps to preclude isomorphism from existing. The structure for a quantum Fourier transform is shown in Fig.~\ref{fig:genericfouriertransform}.

The final primitive, $\mathfrak{U}_{\phi}(\theta)$ corresponds to a diagonal phasing operation on the various qubits corresponding to the eigenvalues of the various irreps of the kinetic term of the Hamiltonian. 

\section{Inversion Gates}
\label{sec:inverse}

\begin{figure}[!ht]
\includegraphics[width=\linewidth]{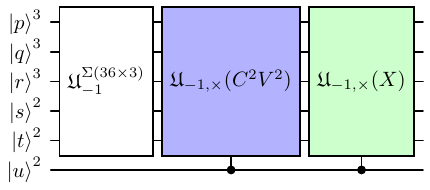}\caption{Diagrammatic formulation of $\mathfrak{U}_{-1}^{\Sigma(72\times3)}$ with $\mathfrak{U}_{-1}^{\dsd}$ from~\cite{Gustafson:2024kym} and the remaining gates defined in Fig.~\ref{fig:inversedecomp} and~\ref{fig:qubitdsainverse}. }
\label{fig:diagramaticinverse}
\end{figure}

\begin{figure}[ht]
    \centering
    
\includegraphics[width=0.6\linewidth]{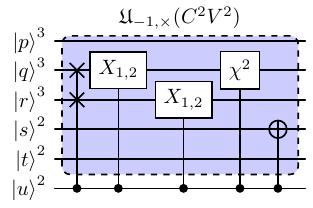}\vspace{0.2cm}
\includegraphics[width=\linewidth]{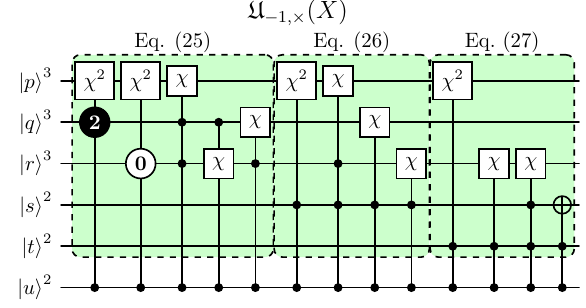}
    \caption{Decomposition for (top) $\mathfrak{U}_{-1,\times}(C^2V^2)$ and (bottom) $\mathfrak{U}_{-1,\times}(X)$ (right) into qubit-qutrit gates which appear in the circuit decomposition of Fig. \ref{fig:diagramaticinverse}.}
    \label{fig:inversedecomp}
\end{figure}

The inversion operations $\mathfrak{U}_{-1}^{\Sigma(72\times3)}$ can be written in a structured recursive way as shown in Fig. \ref{fig:diagramaticinverse} which relies on the inversion gate of $\dsd$. To derive this, first one establishes the inversion relation rules for the bit-strings $pqrstu$. This starts by identifying what happens to the group element $g \in \dsa$ in Eq.~(\ref{eq:s216orderedproduct}) under inversion:
\begin{align}
\label{eq:propinv0}
    g^{-1} = &X^{-u} V^{2(s + t) + t} E^{2r} C^{2q} \omega^{2p} \notag\\
    =  &X^{u} C^{2u} V^{2u} (\omega^{p'}C^{q'}E^{r'}V^{2s'+t'})
\end{align}
where $p'$, $q'$, $r'$, $s'$, and $t'$ are after application of $\mathfrak{U}_{-1}^{\dsd}$  found in~\cite{Gustafson:2024kym}.
The next step, which we will implement as the gate $\mathfrak{U}_{\times}(C^2V^2)$, involves propagating the $C^{2u}V^{2u}$ through the remainder of the expression to get:
\begin{align}
    g^{-1} = X^{u} \omega^{p''} C^{q''} E^{r''} V^{2s'' + t''}
\end{align}
where since $\omega$ commutes and $V^{2u}$ is squared $p''=p'$ and $t''=t'$ are unaffected leaving:
\begin{align}
    q'' = & q'(1-u) + 2 u (q' + 1)\notag\\
    r'' = & r'(1-u) + 2 u r'\notag\\
    s'' = & (s' + u)
\end{align}
To proceed further, one pushes the $X^u$ through the remaining elements. The associated gate we call $\mathfrak{U}_{\times}(X)$ and has three steps. Propagating $X^u$ first through the $C^{q''}E^{r''}$ terms yields $g^{-1} = \omega^{p'''} C^{q'''} E^{r'''} X^u V^{2s''' + t'''}$ with
\begin{align}
\label{eq:propinv2}
    p''' = & p'' + u(2\delta_{q',2} + 2\delta_{r,1} + 2\delta_{r',2} + q'r')\notag\\
    q''' = & q''(1-u) + (2q'' + r'')u\notag\\
    r''' = & r'' + q''u
\end{align}
with $s'''=s'',t'''=t''$ unchanged. Going through $V^{2s}$ gives $g^{-1}=\omega^{\tilde{p}}C^{\tilde{q}}E^{\tilde{r}}V^{2\tilde{s}}X^uV^{\tilde{t}}$
\begin{align}
    \tilde{p} = & p''' + 2  u  s''' +  u s''' r''' \notag\\
    \tilde{q} = & q''' + s'''u\notag\\
    \tilde{r} = & r''' + s'''u\notag\\
    \tilde{s} = & s'''
\end{align}
while leaving $\tilde{t}=t'''$ untouched. The final push through $V^t$ give $g^{-1} = \omega^{\bar{p}}C^{\bar{q}}E^{\bar{r}}V^{2\bar{s} + \bar{t}}X^{\bar{u}}$ which interestingly changed neither $\bar{t}=\tilde{t}$ nor $\bar{u}=u$ but instead:
\begin{align}
\label{eq:propinv4}
    \bar{p} = & \tilde{p} + 2u\tilde{t}\notag\\
    \bar{q} = & \tilde{q}\notag\\
    \bar{r} = & \tilde{r} + u\tilde{t}(1-\tilde{s}) + 2u\tilde{t}\tilde{s}\notag\\
    \bar{s} = & \tilde{s} + u\tilde{t}
\end{align}
The explicit circuit which performs $\mathfrak{U}_{\times}(C^2V^2)$ and $\mathfrak{U}_{\times}(X)$ are provided in Fig. \ref{fig:inversedecomp}. One can convert this qutrit-qubit into the purely qubit gate shown in Fig.~\ref{fig:qubitdsainverse} using the methods of Appendix~\ref{app:toffolistaircase}.

\begin{figure*}
\includegraphics[width=0.22\linewidth]{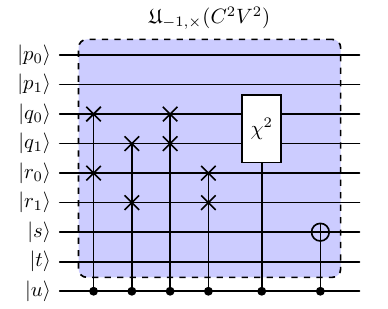}
\includegraphics[width=0.75\linewidth]{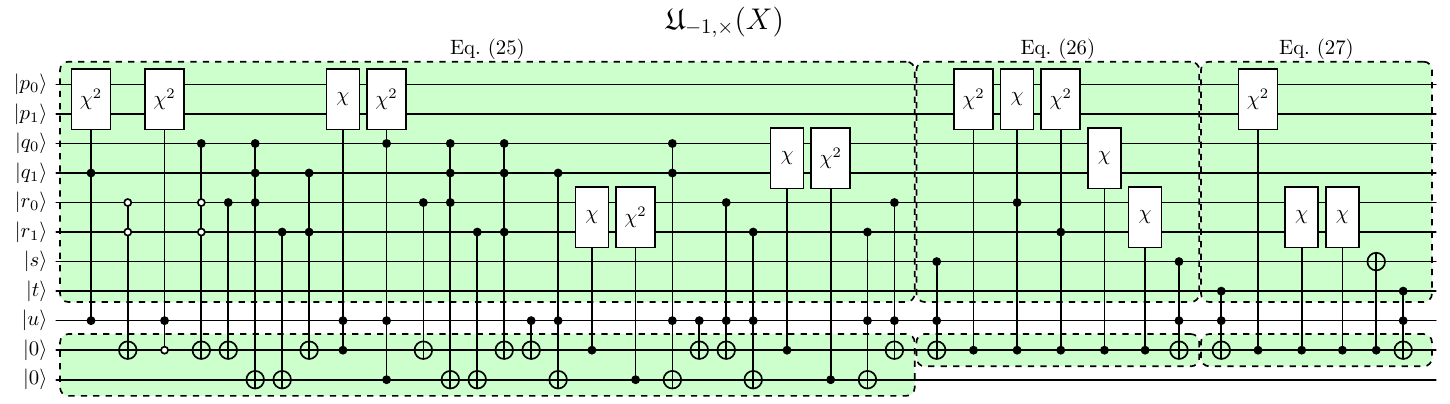}
\caption{Decomposition of the intermediate controlled inversion gates (left) $\mathfrak{U}_{-1,\times}(C^2V^2)$ and (right) $\mathfrak{U}_{-1,\times}(X)$. The control is explicitly on the $|u\rangle$ state and the target qubits are ancilla registers and the $\Sigma(36\times3)$ subgroup.}
\label{fig:qubitdsainverse}
\end{figure*}

\section{Multiplication Gates}
\label{sec:multiplication}

The group multiplication gate uses a similar recursive decomposition to $\mathfrak{U}_{-1}$ by breaking it into two gates:
\begin{align}
    \mathfrak{U}_{\times}^{\dsa} = \mathfrak{U}_{\times}^{\Sigma(36\times3)} \mathfrak{U}_{\times, X}.
\end{align}
The circuit for $\mathfrak{U}_{\times}^{\Sigma(36\times3)}$ is found in~\cite{Gustafson:2024kym}. The additional $\mathfrak{U}_{\times, X}$ are modifications gates which appear earlier in the inversion procedure. To start, given group elements
\begin{align*}
    g =& \omega^{p_g}C^{q_g}E^{r_g}V^{2s_g + t_g}X^{u_g}\notag\\
    h = &\omega^{p_h}C^{q_h}E^{r_h}V^{2s_h+t_h}X^{u_h}\notag
\end{align*}
the product of the two elements can be written as
\begin{align}
    hg = \omega^{\bar{p}}C^{\bar{q}}E^{\bar{r}}V^{2\bar{s}+\bar{t}}X^{\bar{u}}
\end{align}
where we need to determine the exponents. The first step propagates $X^{u'}$ yielding the same $\mathfrak{U}_{-1,\times}(X)$ except controlled on $\ket{u_h}$. 
For the case of $u_g=2$, the application of $\mathfrak{U}_{-1,\times}(X)$ alone is insufficient in the case where $u_h=u_g'=1$, as it produces an additional $C^2V^2$:
\begin{equation}
        hg = \omega^{p_g'}C^{q_g'}E^{r_g'}V^{2s_g'+t_g'}C^{2u_hu_g}V^{2u_hu_g}X^{u_g\oplus u_h}
\end{equation}
which must then be propagated into place, giving
\begin{align}
    p_g''=&p_g'\oplus_3r_g'u_gu_h(1-t_g')(s_g' \oplus_3 2(1 - s_g'))\notag\\
    q_g''=&q_g' \oplus_3 u_gu_h(1-t_g')(s_g' \oplus_3 2 (1-s_g'))\notag\\
    r_g''=&r_g' \oplus_3 u_gu_ht_g'(2(1-s_g') \oplus_3 s_g')\notag\\
    s_g''=&s_g'\oplus_2 su_gu_h\notag\\
    t_g''=&t_g'\notag\\
    u_g''=&u_g \oplus_2 s u_h
\end{align}

\begin{figure}[!ht]
    \centering
\includegraphics[width=\linewidth]{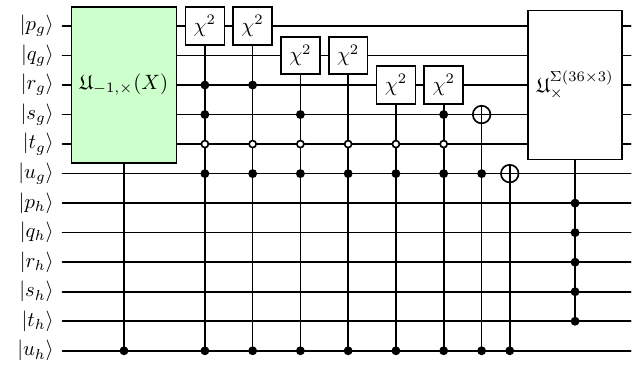}
    \caption{Qutrit-Qubit decomposition of  $\mathfrak{U}_{\times}^{\dsa}$. $\mathfrak{U}_{-1,\times}(X)$ is shown in Fig. \ref{fig:diagramaticinverse} except the control is $|u_h\rangle$ instead of $|u_g\rangle$ while $\mathfrak{U}_{\times}^{ \Sigma(36\times3)}$ is provided in Ref. \cite{Gustafson:2024kym}.}
    \label{fig:dsamultiply}
\end{figure}

\begin{figure}[!ht]
\centering
    \includegraphics[width=\linewidth]{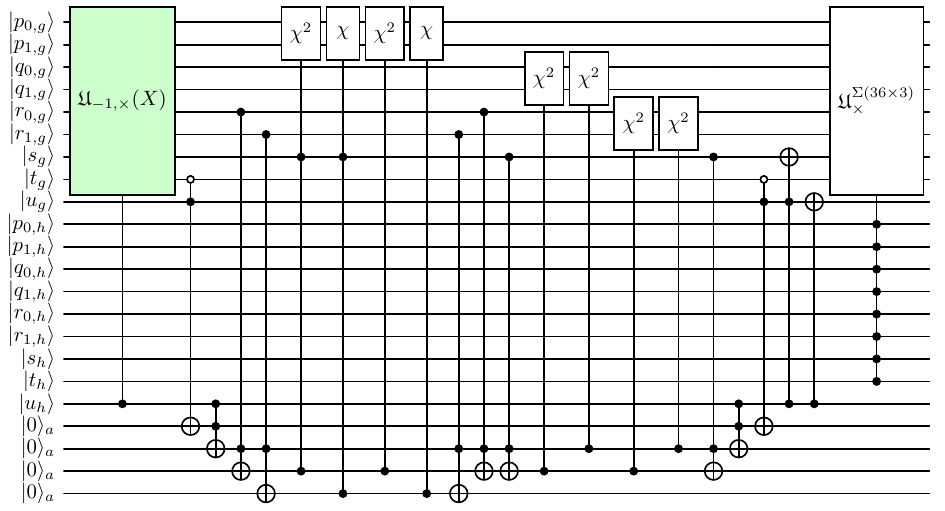}
\caption{Qubit decomposition of $\mathfrak{U}_{\times}^{\dsa}$. $\mathfrak{U}_{-1,\times}(X)$ is shown in Fig. \ref{fig:qubitdsainverse} except the control is $|u_h\rangle$ instead of $|u_g\rangle$.}
\label{fig:qubitmultiplicationdsa.}
\end{figure}

After this, the remaining operations are equivalent to $\mathfrak{U}_{\times}^{\dsd}$ which can be performed following~\cite{Gustafson:2024kym}.
The qutrit-qubit circuit is shown in Fig.~\ref{fig:dsamultiply}, while the qubit circuit is provided in Fig.~\ref{fig:qubitmultiplicationdsa.}.

\section{Trace Gates}
\label{sec:trace}

\begin{figure}[!ht]
\includegraphics[width=0.8\linewidth]{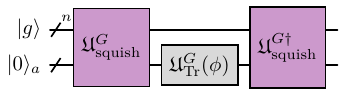}
\caption{Schematic of  $\mathfrak{U}_{\text{Tr}}$ for an arbitrary group $G$. The $\mathfrak{U}_{\rm{squish}}^{G}$ maps the group elements to the real values of the trace classes and the gate $\mathfrak{U}_{\rm{Tr}}^{G}(\phi)$ phases the group elements depending on their real values of the fiducial trace.}
\label{fig:traceexample}
\end{figure}

The trace gate $\mathfrak{U}_{\rm{Tr}}$ for a discrete group can be decomposed into two components: $\mathfrak{U}_{\rm{squish}}^{ G}$ and 
$\mathfrak{U}_{\rm{Tr}}^{ G}(\phi)$.
The circuits $\mathfrak{U}_{\rm{squish}}^{G}$ and its conjugate compute and uncompute the associated trace class for $g$ onto an ancilla register. With this identification, $\mathfrak{U}_{\rm{Tr}}^{G}$ implements a small Hamiltonian on the ancilla register to get the correct phase rotation. A schematic for $\mathfrak{U}_{\rm{Tr}}$ is provided in Fig.~\ref{fig:traceexample}.

To construct $\mathfrak{U}_{\rm{Tr}}$, we need to construct a map from $g$ to their respective trace classes found in Tab.~\ref{tab:cts216}. For $\dsa$, the real portion of the trace for the classes are identical to those of $\Sigma(36\times 3)$: $\lbrace 3, - \frac{3}{2}, 0, \pm 1, \pm\frac{1}{2}\rbrace$ and how each trace is mapped into the ancilla is in Tab. \ref{tab:tracedsa}.

\begin{table}[!th]
\caption{Map for the group elements to ancilla register for $\dsa$. This mapping is the same as that for $\Sigma(36\times3)$. }
\label{tab:tracedsa}
\begin{tabular}{c|ccccccc}
\hline\hline
$\rm{Re}(\rm{Tr}(g))$ &$3$&$-\frac{3}{2}$&$-1$&$-\frac{1}{2}$&$0$&$1$&$\frac{1}{2}$\\
\hline 
$v_0$ & 0 & 1 & 0 & 1 & 0 & 1 & 0\\
$v_1$ & 0 & 0 & 1 & 1 & 0 & 0 & 1\\
$v_2$ & 0 & 0 & 0 & 0 & 1 & 1 & 1\\\hline\hline
\end{tabular}
\end{table}

\begin{figure}[!ht]
\includegraphics[width=\linewidth]{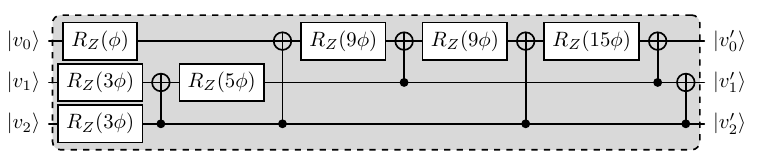}
\caption{Implementation of $\mathfrak{U}_{\text{Tr}}^{\dsa}(\phi)$.}
\label{fig:dsatraceham}
\end{figure}

Using Tab. \ref{tab:tracedsa}, the $\mathfrak{U}_{\text{Tr}}^{\dsa}(\phi)$ is shown in Fig. \ref{fig:dsatraceham} and is same as the one for $\Sigma(36\times3)$ from~\cite{Gustafson:2024kym}.

\begin{figure}[!ht]
\includegraphics[width=\linewidth]{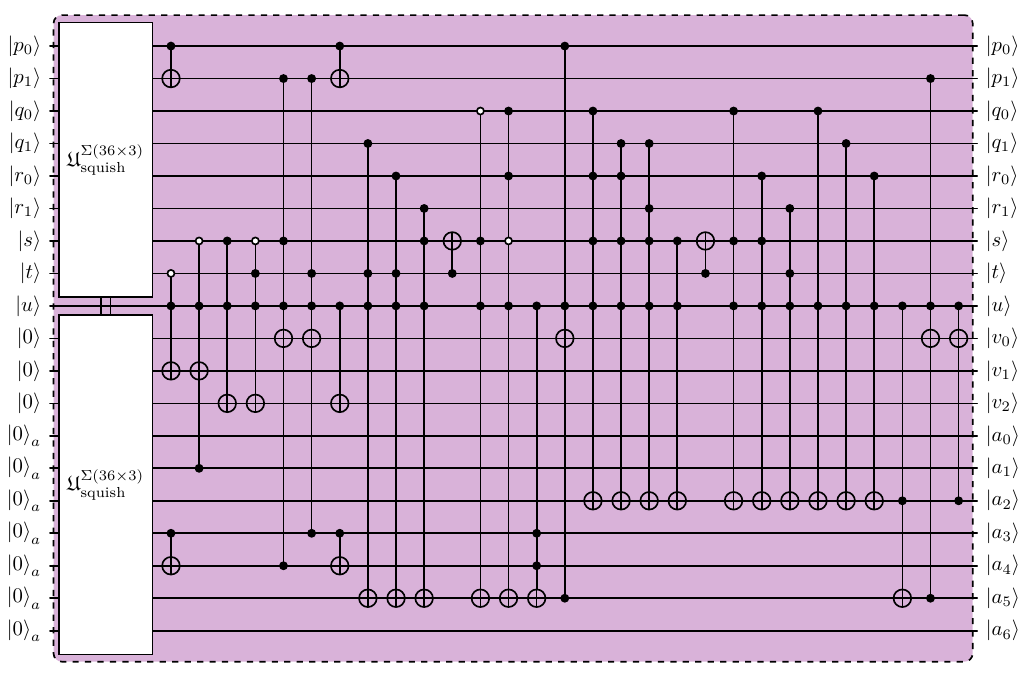}
\caption{Circuit for $\mathfrak{U}_{\rm squish}^{\dsa}$ using $\mathfrak{U}_{\rm squish}^{\dsd}$ from~\cite{Gustafson:2024kym}.}
\label{fig:squish216}
\end{figure}

The difference with $\dsa$ is that there are additional conjugacy classes that need to be mapped to $\ket{v_2v_1v_0}$. The extra conjugacy classes we need are:
\begin{align}
\label{eq:conclass}
        C_{X} = \{& X, \omega^2V^2X, \omega^2E^2V^2X, \omega^2C^2EV^2X, \omega C^2X, \notag\\
        &\omega EV^2X, C^2V^2X, \omega CX, \omega^2CV^2X, \omega^2CE^2V^2X, \notag\\
        &\omega EX, \omega CEV^2X, \omega C^2EX, C^2E^2V^2X, \notag\\
        &CE^2X, \omega^2C^2E^2X, CEX, E^2X \},\notag\\
        C_{VX} = \{& VX, CV^3X, \omega CEV^3X, \omega^2 C^2EV^3X, \omega^2EVX, \notag\\
        &\omega^2C^2VX, CE^2V^3X, \omega C^2E^2V^3X, \omega^2CEVX,\notag\\ &\omega CE^2VX,
        \omega E^2V^3X, \omega^2CVX, C^2EVX, E^2VX, \notag\\
        &\omega C^2V^3X, \omega V^3X, C^2E^2VX, \omega^2EV^3X \},\notag
\end{align}
as well as $\omega C_{X}$, $\omega^2 C_{X}$, $\omega C_{VX}$, and $\omega^2 C_{VX}$. We thus need to include additional maps for these classes:
\begin{align}
    \re(\tr(g)) \mapsto \begin{cases}
        - \frac{1}{2} & g \in C_{VX},\,\omega C_{VX},\,\omega C_X,\,\omega^2 C_{X}\\
        1 & g \in C_X,\,\omega^2 C_{VX}
    \end{cases}.
\end{align}
Working through the integer logic for each of the elements in Eq.~(\ref{eq:conclass}) leads to the additional circuit structure seen in Fig.~\ref{fig:squish216} for $\mathfrak{U}_{\rm squish}^{\dsa}$.

\section{Fourier Transformations}
\label{sec:fourier}

The electric field Hamiltonian for a group $G$ can be written
\begin{align}
\label{eq:electric}
    \hat{H}_{E} =& \sum_{h \in \Gamma}\sum_{g\in G} |hg\rangle\langle g|\notag\\
    = & \sum_{\rho,m,n}f(\rho)|\rho, m, n\rangle\langle \rho, m, n|,
\end{align}
where $f(\rho) = \vert \Gamma \vert - \frac{1}{\text{dim}(\rho)}\sum_{h\in \Gamma}\tr(\rho(h))$, and $\Gamma$ is a subset of $G$ subject to the constraints:
\begin{align}
    \Gamma^{-1} = & \Gamma \notag\\
    g \Gamma g^{-1} = &\Gamma ~\forall g\in G\notag\\
    \mathbbm{1} \notin & ~\Gamma,
\end{align}
guaranteeing gauge-invariance. While the choice of $\Gamma$ is not unique, the particular choice
\begin{align}
    \Gamma = &\{g \in G\backslash\{1\} \; | \; {\rm Re Tr}(g) \; {\rm is \; maximal}\,\}
\end{align}
can be derived from the Wilson action via the transfer matrix procedure~\cite{Harlow:2018tng,PhysRevD.107.114513}.

Since the electric term is diagonal in the basis labeled by the irreducible representations, we will employ the Fourier transformation to move between the position basis $\{\ket{g}\}$ and the $\{\ket{\rho, m, n}\}$ basis. To this end, we will need to construct the Fourier transform quantum circuit of this group. 

\begin{figure}[!ht]    
\includegraphics[width=0.9\linewidth]{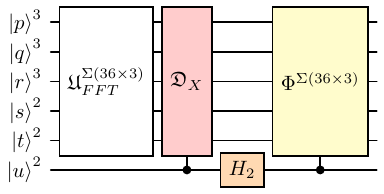}
\caption{Block schematic decomposition of the fast Fourier transformation for $\dsa$.}
\label{fig:fft_schematic}
\end{figure}

The construction of $\mathfrak{U}_{FT}$ for  $\dsa$ follows those in in~\cite{Murairi:2024xpc,1998quant.ph..7064P} and can be written in a block format shown in Fig. \ref{fig:fft_schematic}. The construction of a FFT relies on being able to build the twiddle matrices $\mathfrak{D}_{X}$ and the phase kickback matrices $\Phi^{\Sigma(36\times3)}$.

\begin{figure}[!th]
\includegraphics[width=\linewidth]{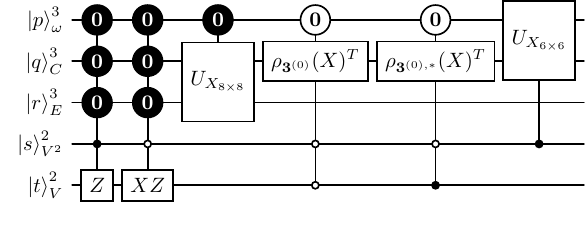}
\caption{Qutrit-qubit implementation of $\mathfrak{D}_{X}$ for  $\dsa$.}
\label{fig:Dx}
\end{figure}

\begin{figure}[!ht]
\includegraphics[width=\linewidth]{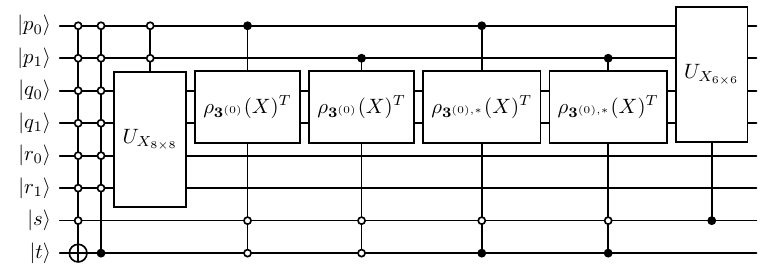}
\caption{Qubit implementation of $\mathfrak{D}_{X}$ for  $\dsa$.}
\label{fig:Dxqubits}
\end{figure}

The first step in building $\mathfrak{D}_{X}$ is identifying which irreps are extendable, i.e. $\xi_{\dsd}(X^{-1} g X) = \xi_{\dsd}(g)$, and which ones for inner conjugates. 
These irreps are:
\begin{align}
    \xi_{\mathbf{1}^{(a)}}(j) = i^{a(2s + t)}&&0\leq a \leq 3,
\end{align}

\begin{align}
    \xi_{\mathbf{3}^{(a, b)}}(j) = (-1)^{bt}\omega^{(1+b)p}C^{(1+b)q}E^r(i^aV)^{2s + t},
\end{align}
where $0\leq a\leq 3$ and $0\leq b\leq 1$, and
\begin{align}
    &\xi_{\mathbf{4}^b}(\omega) = 1,~~ \xi_{\mathbf{4}^b}(C) = \text{Diag}(\omega^b, \omega, \omega^{2b}, \omega^2)\notag\\
    &\xi_{\mathbf{4}^b}(E) = \text{Diag}(\omega, \omega^{2b}, \omega^2, \omega^b)\notag\\
    &\xi_{\mathbf{4}^b}(V) = \begin{pmatrix}
        0 & 1 & 0 & 0\\
        0 & 0 & 1 & 0\\
        0 & 0 & 0 & 1\\
        1 & 0 & 0 & 0
    \end{pmatrix}.
\end{align}
We note that multiplying by $X$ forces $s\rightarrow s + t$, meaning that the trivial irrep, $\xi_{\mathbf{1}^{(0)}}$, and the parity irrep $\xi_{\mathbf{1}^{(2)}}$ are extendable, while the other 1d irreps, $\xi_{\mathbf{1}^{(1)}}$ and $\xi_{\mathbf{1}^{(3)}}$, are a pair of inner conjugates.
The 4d irreps are also a pair of inner conjugates. Similarly, the $\xi_{\mathbf{3}^{(0,b)}}$, $\xi_{\mathbf{3}^{(2,b)}}$ are extendable while, 
$(\xi_{\mathbf{3}^{(1,b)}},\xi_{\mathbf{3}^{(3,b)}})$, are pairs of inner conjugates. 
In this way $\mathfrak{D}_X$ induces $\mathbf{2}^{(0)}$ by mixing the irreps $\xi_{\mathbf{1}^{(1)}}$ and $\xi_{\mathbf{1}^{(3)}}$. The 3d inner conjugates further induce the $\mathbf{6}^{(0)}$ while the 4d irreps induce $\mathbf{8}$.
With this, $\mathfrak{D}_X$ is constructed following~\cite{Murairi:2024xpc}. $\mathfrak{U}_{FFT}^{\Sigma(36\times3)}$ takes the group element basis states to the irrep basis where states are identified as follows:

\begin{align}
    |0\rangle_V^2|0\rangle_{V^2}^2|0\rangle_{E}^3|0\rangle_{C}^3|0\rangle_{\omega}^3 = &|\xi_{\mathbf{1}^{(0)}}\rangle\notag\\
    |1\rangle_V^2|0\rangle_{V^2}^2|0\rangle_{E}^3|0\rangle_{C}^3|0\rangle_{\omega}^3 = &|\xi_{\mathbf{1}^{(2)}}\rangle\notag\\
    |0\rangle_V^2|1\rangle_{V^2}^2|0\rangle_{E}^3|0\rangle_{C}^3|0\rangle_{\omega}^3 = &|\xi_{\mathbf{1}^{(1)}}\rangle\notag\\
    |1\rangle_V^2|1\rangle_{V^2}^2|0\rangle_{E}^3|0\rangle_{C}^3|0\rangle_{\omega}^3 = &|\xi_{\mathbf{1}^{(3)}}\rangle\notag\\
    \lbrace |0\rangle_V^2|a\rangle_{V^2}^2|b\rangle_{E}^3|c\rangle_{C}^3|0\rangle_{\omega}^3 ~|~ b=c \neq 0\rbrace = &|\xi_{\mathbf{4}^{(0)}}\rangle\notag\\
    \lbrace |1\rangle_V^2|a\rangle_{V^2}^2|b\rangle_{E}^3|c\rangle_{C}^3|0\rangle_{\omega}^3 ~|~b=c\neq 0\rbrace = &|\xi_{\mathbf{4}^{(1)}}\rangle\notag\\
    |0\rangle_V^2|0\rangle_{V^2}^2|a\rangle_{E}^3|b\rangle_{C}^3|1\rangle_{\omega}^3 = &|\xi_{\mathbf{3}^{(0, 0)}}\rangle\notag\\
    |1\rangle_V^2|0\rangle_{V^2}^2|a\rangle_{E}^3|b\rangle_{C}^3|1\rangle_{\omega}^3 = &|\xi_{\mathbf{1}^{(0, 1)}}\rangle\notag\\
    |0\rangle_V^2|1\rangle_{V^2}^2|a\rangle_{E}^3|b\rangle_{C}^3|1\rangle_{\omega}^3 = &|\xi_{\mathbf{3}^{(2,0 )}}\rangle\notag\\
    |1\rangle_V^2|1\rangle_{V^2}^2|a\rangle_{E}^3|b\rangle_{C}^3|1\rangle_{\omega}^3 = &|\xi_{\mathbf{3}^{(2,1)}}\rangle\notag\\
    |0\rangle_V^2|0\rangle_{V^2}^2|a\rangle_{E}^3|b\rangle_{C}^3|2\rangle_{\omega}^3 = &|\xi_{\mathbf{3}^{(1, 0)}}\rangle\notag\\
    |1\rangle_V^2|0\rangle_{V^2}^2|a\rangle_{E}^3|b\rangle_{C}^3|2\rangle_{\omega}^3 = &|\xi_{\mathbf{1}^{(1, 1)}}\rangle\notag\\
    |0\rangle_V^2|1\rangle_{V^2}^2|a\rangle_{E}^3|b\rangle_{C}^3|2\rangle_{\omega}^3 = &|\xi_{\mathbf{3}^{(3,0 )}}\rangle\notag\\
    |1\rangle_V^2|1\rangle_{V^2}^2|a\rangle_{E}^3|b\rangle_{C}^3|2\rangle_{\omega}^3 = &|\xi_{\mathbf{3}^{(3,1)}}\rangle.
\end{align}
With this identification one writes $\mathfrak{D}_X$: 
\begin{align}
    \langle 00000|\mathfrak{D}_X|00000\rangle = & 1\notag\\
    \langle 10000|\mathfrak{D}_X|10000\rangle = & -1\notag\\
    \langle t1000|\mathfrak{D}_X|t'1000\rangle = &\Big(\rho_{\mathbf{2}}(X)\Big)_{t,t'}\notag\\
    \langle tsrq0|\mathfrak{D}_X|tsr'q'0\rangle = &\Big(\rho_{\mathbf{8}}(X)\Big)_{3q + r - 1, 3q' + r'-1}\notag\\
    ~q=r\neq0;~q'=r'\neq0&\notag\\
    \langle 00rqp|\mathfrak{D}_{X}|00rq'p\rangle = & \Big(\rho_{\mathbf{3}^{(0)}}(X)\Big)^T_{q, q'}; ~p \in \lbrace 1, 2\rbrace & \notag\\
    \langle 10rqp|\mathfrak{D}_{X}|10rq'p\rangle = & \Big(\rho_{\mathbf{3}^{(0),*}}(X)\Big)^T_{q, q'}; ~p \in \lbrace 1, 2\rbrace & \notag\\
    \langle t10qp|\mathfrak{D}_X|t10q'p'\rangle = & \Big(\rho_{\mathbf{6}}(X)\Big)_{3(p-1) + q,3(p'-1)+q'}\notag\\
    p\neq0,~p'\neq0&
\end{align}
The quantum circuit for $\mathfrak{D}_X$ is provided in Fig. \ref{fig:Dx} with
\begin{align}
    U_{X_{8\times8}} &= 1\oplus \rho_{\mathbf{8}}(X)\notag\\
    U_{X_{6\times6}} &= \mathbbm{1}_3 \oplus \rho_{\mathbf{6}}(X).
\end{align}
$U_{X_{8\times8}}$ and $U_{X_{6\times6}}$ can be decomposed using the quantum Shannon decomposition (QSD) (Refs.~\cite{PAIGE1994303,CHEN2013853,PhysRevA.92.062317,Gustafson:2024kym}). As shown in Fig.~\ref{fig:Dx}, we need to implement the controlled C$U_{X_{8\times8}}$ as well as C$U_{X_{6\times6}}$ while minimizing non-Clifford gates. This can be done via ZX calculus~\cite{Duncan2020graphtheoretic} where we use \verb|pyzx|~\cite{kissinger2020Pyzx}. For C$U_{X_{8\times8}}$, we obtain 156 $R_Z$ and 301 Clifford gates. On the other hand, we have 158 $R_Z$ and 215 Clifford gates for C$U_{X_{6\times6}}$. Finally, Fig.~\ref{fig:Dxqubits} shows the qubit decomposition of the circuit in Fig~\ref{fig:Dx}. To obtain the final qubit gate cost of this circuit, it remains to decompose the two C$\rho_{\mathbf{3}^{(0)}}(X)^T$ and C$\rho_{\mathbf{3}^{(0),*}}(X)^T$. These four controlled gates reduce to only two controlled gates with the use of two ancillas and four CCNOT gates resulting in a total cost for C$\rho_{\mathbf{3}^{(0)}}(X)^T$ and C$\rho_{\mathbf{3}^{(0),*}}(X)^T$ of 53 $R_Z$ gates, $4$ CCNOT and 2 ancillas.

Having constructed the Fourier gate, one need only implement a set of rotations along the diagonal of the transformed register, with angles given by the eigenvalues of the electric field Hamiltonian  rescaled by trotter step and couplings.  The eigenvalues can be found in Table~\ref{tab:eham}.

\begin{table}[!ht]
\caption{Eigenvalues of the $\dsa$ electric field Hamiltonian normalized by $g^2/2$. We combine certain irreps which have the same eigenvalue by denoting algebraic superscript $a$ taking the values $1\leq a\leq 3$.}
\label{tab:eham}
\begin{tabular}{c|cccccccccc}
\hline\hline
$\rho$ & $\mathbf{1}^{(0)}$ & $\mathbf{1}^{(a)}$ & $\mathbf{2}^{(0)}$ & $\mathbf{3}^{(0)}$& $\mathbf{3}^{(a)}$&$\mathbf{3}^{(0),*}$& $\mathbf{3}^{(a),*}$&$\mathbf{6}^{(0)}$ & $\mathbf{6}^{(1)}$ & $\mathbf{8}$ \\\hline
$f(\rho)$ & 0 & 72 & 54 & 36 & 60 & 36 & 60 & 54 & 54 & 54 \\\hline\hline
\end{tabular}
\end{table}

\section{Resource Costs}
\label{sec:resources}

In order to estimate the fault-tolerant cost of our digitization, we use the following conversions.
The Toffoli gate requires 7 T gates~\cite{Chuang:1996hw} and C$^n$NOT gates can be constructed from $4(n-1)$ Toffoli gates and $n-1$ clean, reusable ancilla qubits\footnote{A \emph{clean} ancilla is a qubit initialized to $\ket{0}$. \emph{Dirty} ancilla indicate ones in an unknown initial state.}~\cite{Chuang:1996hw,PhysRevA.52.3457}. We arrive at the cost for the $R_Z$ gates via \cite{PhysRevLett.114.080502} where these gates can be approximated to precision $\epsilon$ with on average $1.15 \log(1/\epsilon)$ T gates~\cite{PhysRevLett.114.080502}. Additionally, $R_Y$ and $R_X$ can be replaced by at most 1 $R_Z$ gate and 2 Clifford transformations. Using these, we can construct fault-tolerant estimates assuming the T gate dominates, although this assumption has become questionable~\cite{Gidney:2024alh}. The number of T gates necessary to implement the various basic and primitive gates can be found in Tab. \ref{tab:tgatecost}.

\begin{table}[]
    \centering
    \caption{T gate counts and clean ancilla required to implement gates from~\cite{Chuang:1996hw}}
    \label{tab:tgatecost}
    \begin{tabular}{ccc}
    \hline\hline
         Gate & T gates & Clean ancilla\\
         \hline
         C$^2$NOT & 7 & 0\\
         C$^3$NOT & 21 & 1\\
         C$^4$NOT & 35 & 2\\
         CSWAP & 7 & 0\\
         $R_Z$ & 1.15 $\log_2(1/\epsilon)$ & 0\\\hline
         $\mathfrak{U}_{-1}$ &637&4 \\
         $\mathfrak{U}_{\times}$&1120&5\\
         $\mathfrak{U}_{\text{Tr}}$ & 1414+8.05 log$_2(1/\epsilon)$& 12\\
         $\mathfrak{U}_{FT}$&186295.4 log$_2(1/\epsilon)$ + 112& 5\\
         $\mathfrak{U}_{\phi}$&294.4log$_2(1/\epsilon)$&0\\\hline\hline
    \end{tabular}
\end{table}

\begin{figure*}
\includegraphics[width=\linewidth]{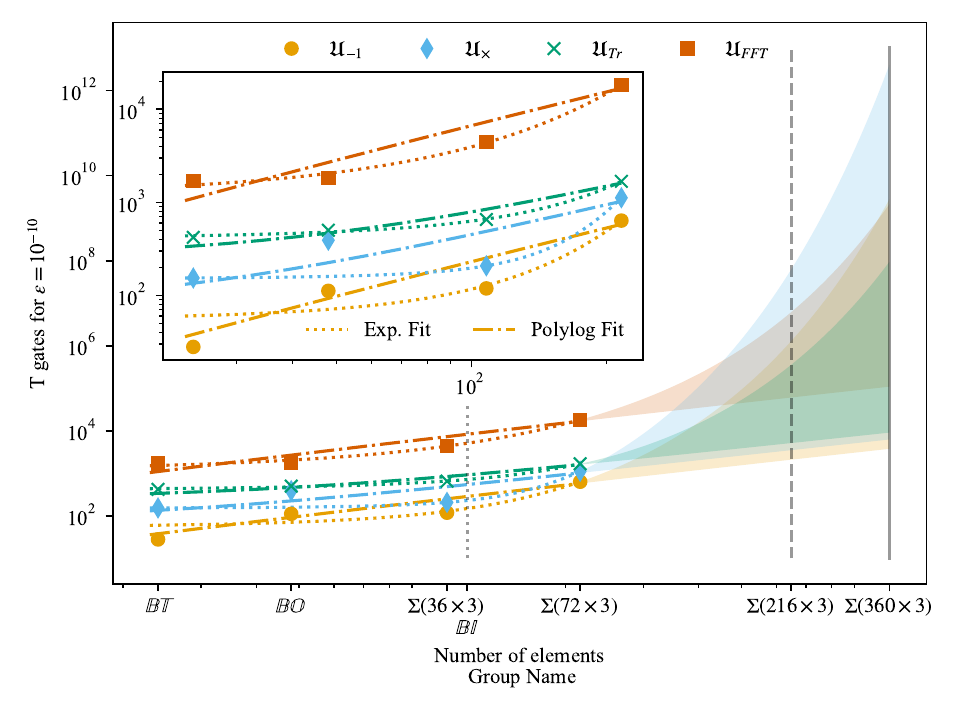}
\vspace{-2em}
\caption{T gate costs for the primitive gates $\mathfrak{U}_{-1}$, $\mathfrak{U}_{\times}$, $\mathfrak{U}_{Tr}$, and $\mathfrak{U}_{FFT}$ for the fixed synthesis precision $\epsilon=10^{-10}$. Two extrapolations are shown: exponential and linear-logarithmic functional.}
\label{fig:resourcecosts}
    
\end{figure*}

One can consider combining these results with those of existing discrete subgroup primitives to estimate the scaling of these gates with $|G|$.  A mild obstacle to this is we have only determined $\mathfrak{U}_{FT}$ for $\dsa$ rather than the more efficient $\mathfrak{U}_{FFT}$ which are known for the other groups. To remedy this, we attempt the following estimate for a lower bound on the cost of $\mathfrak{U}_{FFT}^{\dsa}$: replace the $\mathfrak{U}_{FT}^{\Sigma(36\times3)}$ used in $\mathfrak{U}_{FT}^{\dsa}$ by the known $\mathfrak{U}_{FFT}^{\Sigma(36\times3)}$. Under this assumption, the cost for $\mathfrak{U}_{FFT}^{\dsa}$ is estimated to be $515.2  \text{log}_2(1/\epsilon) + 532$ T-gates, a 3 order of magnitude reduction compared to the $\mathfrak{U}_{FT}$ . This is a strict lower bound on the cost of $\mathfrak{U}_{FFT}^{\dsa}$ since this replacement can always be done at the cost of a more complex $\mathfrak{D}_X$. In Fig.~\ref{fig:resourcecosts}, we compare the $T$-gate cost estimates for each primitive gate at a fixed error of $\epsilon=10^{-10}$ for $\dsa$ to $\mathbb{BT}$, $\mathbb{BO}$, $\Sigma(36\times3)$, and $\Sigma(72\times3)$. Across all groups, we find the ordering of costs is the same: $\mathfrak{U}_{-1}$, $\mathfrak{U}_{\times}$, $\mathfrak{U}_{\text{Tr}}$, $\mathfrak{U}_{FFT}$ from cheapest to most expensive.

An interesting result is unlike $\mathbb{BT}$ and $\Sigma(36\times3)$ where $\mathfrak{U}_{\text{Tr}}$ requires $\sim 5\times$ more T gates than $\mathfrak{U}_{\times}$, for $\bo$ and $\dsa$ the two gates has similar cost.  It will be interesting to see if the remaining, larger groups $\bi$, $\dsb$, and $\dsc$ keep this existing hierarchy.  It is conceivable that for these groups that $\mathfrak{U}_{\times}$ becomes more expensive, since the number of trace classes grows sublinearly in $|G|$ despite fact that $\mathfrak{U}_{\text{Tr}}$ requires $R_Z$ synthesis. 

Using our existing results, we can attempt to make extrapolations to predict the costs of remaining larger $SU(3)$ groups. One should be careful with this analysis since the $SU(2)$ and $SU(3)$ subgroups \textit{do not} form a tower of groups.  Thus prefactors in scaling relations may change between different $SU(N)$. Nevertheless using these results, we fit to two functional forms. The first is a linear logarithmic form for a ``lower" bound:
\begin{align}
    f(|G|) = A + B |G| \text{log}_2(|G|),
\end{align}
where $|G|$ is the size of the finite group $|G|$. We denote this fit form as polylog in the figure. The second is an ``upper" bound that goes exponentially in group size:
\begin{align}
    h(|G|) = A + B e^{C |G|}.
\end{align}
The best fits for these forms are displayed alongside the known gate costs in Fig.~\ref{fig:resourcecosts} and Tab. \ref{tab:estimates} contains the numerical predictions. These results suggest that the $\dsb$ and $\dsc$ gates will likely cost between $\sim 10\times$ and  $\sim 10^8\times$  more T gates than $\dsa$. 

\begin{table}[!th]
\caption{Estimated number of T gates for $\dsb$ and $\dsc$.}
\label{tab:estimates}
\begin{tabular}{ccc}
\hline\hline Gate & $\dsb$ & $\Sigma(360\times3)$\\\hline
$\mathfrak{U}_{-1}$ & $10^3-10^6$ &$10^3-10^9$\\
$\mathfrak{U}_{\times}$ & $10^3-10^{7}$ & $10^3-10^{12}$\\
$\mathfrak{U}_{Tr}$ & $10^3-10^7$& $10^5-10^9$\\
$\mathfrak{U}_{FFT}$ & $10^4-10^6$ & $10^5-10^9$\\
\hline\hline 
\end{tabular}
\end{table}

Following \cite{Cohen:2021imf,Kan:2021xfc,Gustafson:2022xdt}, we integrate the primitive gates as subroutines into a quantum algorithm for computing the shear viscosity $\eta$ on a $L^3=10^3$ lattice evolved for $N_t=50$, and total synthesis error of $\epsilon_T=10^{-8}$.  We are neglecting state preparation (which can be substantial~\cite{Xie:2022jgj,Davoudi:2022uzo,Avkhadiev:2019niu,Gustafson:2022hjf,peruzzo2014variational,Gustafson:2019mpk,Gustafson:2019vsd,Harmalkar:2020mpd,Gustafson:2020yfe,Jordan:2017lea,PhysRevLett.108.080402,motta2020determining,Clemente:2020lpr,motta2020determining,deJong:2021wsd,Gustafson:2023ayr,Kane:2023jdo}). We consider a Trotterized time evolution with step size $\delta$ and two Hamiltonians: the Kogut-Susskind  $H_{KS}$~\cite{PhysRevD.11.395} and the Symanzik-improved $H_I$~\cite{Carena:2022kpg}.  For each of these, the number of primitive gates required per link per $\delta t$ are listed in Tab.~\ref{tab:primcost}. 

\begin{table}[ht]
   \caption{Number of primitive gates per link per $\delta t$ neglecting boundary effects as a function of $d$ for $H_{KS}$ and $H_{I}$.}
    \label{tab:primcost}
    \begin{tabular}{c|c|c}
    \hline\hline
    Gate & $N[H_{KS}]$&$N[H_{I}]$\\
    \hline
    $\mathfrak U_F$ & 2  &4\\
    $\mathfrak U_{\rm Tr}$&$\frac{1}{2}(d-1)$ &$\frac{3}{2}(d-1)$\\
    $\mathfrak U_{-1}$& $3(d-1)$ & $2+11(d-1)$\\
    $\mathfrak U_{\times}$& $6(d-1)$ &$4+26(d-1)$\\
    \hline\hline
    \end{tabular}
\end{table}

These results determine the total T gate count $N^{H}_T=C^H_T\times d L^d N_t$ for a $d$ spatial lattice simulated for a time $t=N_t\delta t$. We find that for $H_{KS}$,
\begin{equation}
    C^{KS}_{T}=9338d-9114+(372881+4.025d)\log_2\frac{1}{\epsilon}.
\end{equation}
 We estimate that the total synthesis error $\epsilon_T$ is the sum of $\epsilon$ from each $R_Z$.  In the case of $H_{KS}$ this yields 
\begin{equation}
    \epsilon_T=\frac{1}{2}(648489 + 7 d)d L^d N_t\times \epsilon.
\end{equation}
For $H_I$, the costs increase to
\begin{equation}
    C^{I}_{T}=38248d-32046+(745758+12.075d)\log_2\frac{1}{\epsilon},
\end{equation}
where the total synthesis error is
\begin{equation}
    \epsilon_T =\frac{1}{2}(1296971 + 21 d)d L^d N_t \times\epsilon.
\end{equation}
Kan and Nam estimated $6.5\times10^{48}$ T gates would be required for an pure-gauge $SU(3)$ simulation of $H_{KS}$. This estimate used a truncated electric-field digitization and considerable fixed-point arithmetic. Here, using $\dsa$ to approximate $SU(3)$ requires $7.1\times10^{12}$ T gates for $H_I$ and $3.5\times10^{12}$ T gates for $H_{KS}$.  The T gate density is roughly 1 per $\dsa-$register per clock cycle. Thus $\dsa$ reduces the gate costs of~\cite{Kan:2021xfc} by $10^{35}$.  Comparing to other discrete subgroups, these costs are nearly identical to those for the smaller $\dsd$ without a fast Fourier transform~\cite{Gustafson:2024kym}, but 700 times more expensive than $\dsd$ with the fast Fourier transform~\cite{Murairi:2024xpc}. Similar to the previous results for discrete subgroups of $SU(N)$, without a full fast Fourier transform, $\mathfrak{U}_{F}$ dominates the simulations -- being over 99\% of the computation regardless of Hamiltonian.

While prior comparisons have been carried out with respect to Kan and Nam \cite{Kan:2021xfc}, more recent results have been proposed by Rhodes et al.~\cite{Rhodes:2024zbr} which utilizes block encodings of lattice gauge theories to help accelerate the quantum computation. Using a pure gauge calculation for various ranges of coefficients at the fixed volumes we consider yields roughly $10^{11}-10^{14}$ T gates. 
 For this reason we consider our results to be competitive with Ref. \cite{Rhodes:2024zbr}. It may also prove that the large degree of sparsity present in discrete group approximation gate would yield a moderate improvement in computational costs compared to Trotterization however the systematic uncertainties and renormalization effects would still remain an open question of research although there has been important recent progress~\cite{Kane:2025ybw}.

\section{Outlook}
\label{sec:conc}

In this work, we provided a construction of the quantum primitive gates to simulate the gauge degrees of freedom of $\dsa$ . This enables simulations of $SU(3)$ with higher levels of approximation than previous results based on $\dsd$.
The estimated T-gate cost incurred in computing the shear viscosity using these groups is found to be $7.1\times10^{12}$, a 35 \emph{orders of magnitude} smaller than the T-gate cost estimated in Ref.~\cite{Kan:2021xfc}. Moreover, this cost is on par with~\cite{Rhodes:2024zbr}. Compared to $\dsd$ (where a fast Fourier transform exists), the reduction in digitization error requires increasing the T gates cost by 700$\times$.

With these results, the remaining crystal-like  subgroups of $SU(3)$ are the two largest -- $\dsb$ and $\Sigma(360 \times 3)$ -- which would further reduce digitization errors at the cost of a circuit depth.  Using the scaling observed for previous crystal-like groups, we estimate $\dsb$ may require $10^{13}$ to $10^{17}$ T-gates while $\dsc$ would require between $10^{13}$ to $10^{20}$. These costs would be substantially lower than estimated based on explicit constructions. 

As we look ahead to the completion of the bosonic primitive gates for crystal-like subgroups, the next step is incorporating fermion fields \cite{Florio:2023kel, Zohar:2016iic, Zohar:2018nvl}. Many methods exist to incorporate staggered and Wilson fermions while other fermion discretizations based on Ginsparg-Wilson relations require theoretical developments~\cite{Clancy:2023kla,Singh:2025sye}. It will be valuable to compare the resource costs for using different fermions not only in terms of T-gates but also using other methods such as \cite{Gui:2023coj}.

\begin{acknowledgements}
EG, EM and HL were supported by
the U.S. Department of Energy, Office of Science, National
Quantum Information Science Research Centers,
Superconducting Quantum Materials and Systems Center
(SQMS) under contract number DE-AC02-07CH11359 where EG’s part is under the NASA-DOE interagency agreement SAA2-403602.
SO, HL and EG were supported by the U.S. Department of Energy, Office of Science, Office of High Energy Physics Quantum Information Science Enabled Discovery (QuantISED) programs ``Intersections of QIS and Theoretical Particle Physics" and ``Toward Lattice QCD on Quantum Computers" with EG under award number DE-SC0025940. EG was supported by the NASA Academic Mission Services, Contract No. NNA16BD14C and the Intelligent Systems Research and Development-3 (ISRDS-3) Contract
80ARC020D0010. EG acknowledges support from Universities Space Research Association for final review and edits of draft. This work was produced by Fermi Forward Discovery Group, LLC under Contract No. 89243024CSC000002 with the U.S. Department of Energy, Office of Science, Office of High Energy Physics.

\end{acknowledgements}

\appendix
\section{Staircase decompositions}
\label{app:toffolistaircase}

\input{app_decompcontrolledsum}

\bibliography{ref}

\end{document}

%% file: app_decompcontrolledsum.tex
\begin{figure}[!ht]
    \centering
    \includegraphics[width=0.5\linewidth]{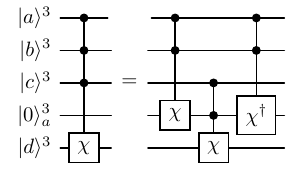}

    \caption{Staircase decomposition of \textsc{CCCSum} into three \textsc{CCSum} with one ancilla.}
    \label{fig:csumstaircase}
\end{figure}

Here, we discuss several decompositions of the $\textsc{CCSum}$ and Toffoli gates. 
The staircase rule for the Toffoli gate can be directly extended to  \textsc{C}$^n$\textsc{Sum} gates regardless of control qudit dimension. One can verify that the decompositions shown in Fig. \ref{fig:csumstaircase} holds. Once can see that the first \textsc{CCSum} in the figure takes the initialized zero ancilla qutrit to the state $|ab\rangle$. The next \textsc{CCSum} uses the control state on the ancilla plus the $|c\rangle$ to shift the bottom most qutrit to the state $|d\oplus_3abc\rangle$. The final \textsc{CCSum} uncomputes the ancilla qutrit. 
\begin{table}[!ht]
\caption{Toffoli and T gate counts for various $\textsc{CCSum}$ gates. The first column corresponds to the dimensions of the control qudit e.g. ``2-3" indicates a qubit and a qutrit control. }
\label{tab:csumcosts}
\begin{tabular}{ccc}
\hline\hline Control & Toffoli gates & T-gates\\\hline
2 & 2 & 14\\
3 & 4 & 28\\
2-2& 4 & 28\\
2-3& 8 & 56\\
3-3& 12 & 84\\
2-2-2& 6& 42\\
3-3-2& 16&112\\\hline\hline
\end{tabular}
\end{table}

\begin{figure}[h]
    \centering

\includegraphics[width=0.54\linewidth]{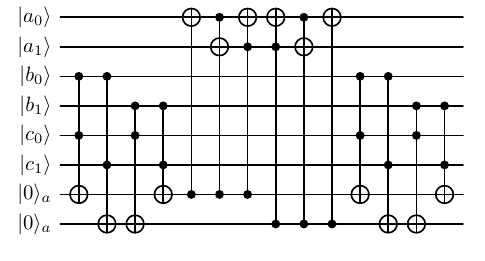}
\includegraphics[width=0.4\linewidth]{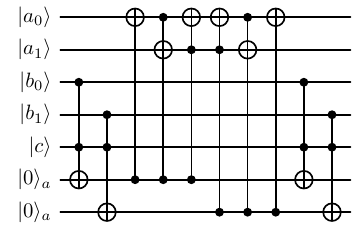}
\includegraphics[width=.6\linewidth]{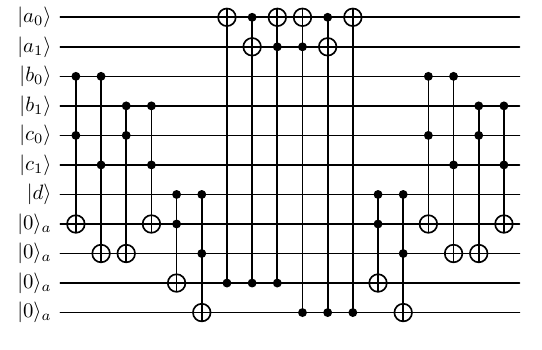}
    \caption{Qubit decompositions of: (top left) a qutrit $\textsc{CCSum}$, (top right) 2-3 $\textsc{CCSum}$ gate,
    (bottom) 2-3-3 $\textsc{CCCSum}$ gate.}
    \label{fig:qubitdecompositioncsum}
\end{figure}

Other decompositions show how the \textsc{C}$^n$\text{Sum} gates are mapped to qubit. Given the qutrit \textsc{CSum} has three possible outcomes ($\mathbbm{1}_3$, $\chi$, and $\chi^2$), one has the circuit in Fig.~ \ref{fig:qubitdecompositioncsum} which applies the correct circuit in the ``physical subspace." The T-gate cost for the 4 Toffolis is 28. One can extend this to \textsc{C}$^n$\text{Sum} as shown in Fig.~\ref{fig:qubitdecompositioncsum}.
In the special case of a \textsc{C}$^n$\text{Sum} controlled by both qubits and qutrits, we end up with a circuit that has fewer required Toffoli gates than in the case with two qutrit registers. Circuits for these and more complicated multicontrolled cases are provided in Fig. \ref{fig:qubitdecompositioncsum}. We provided the T-gate costs and necessary ancilla for decomposing these heterogeneous circuits to homogeneous qubit circuits in Tab. \ref{tab:csumcosts}.

%% file: ref.bib
@article{Aarts_2016,
	Author = {Aarts, Gert},
	Doi = {10.1088/1742-6596/706/2/022004},
	Issn = {1742-6596},
	Journal = {Journal of Physics: Conference Series},
	Month = {Apr},
	Pages = {022004},
	Publisher = {IOP Publishing},
	Title = {Introductory lectures on lattice QCD at nonzero baryon number},
	Url = {http://dx.doi.org/10.1088/1742-6596/706/2/022004},
	Volume = {706},
	Year = {2016}}

@misc{kurkcuoglu2022quantum,
      title={Quantum simulation of $\phi^4$ theories in qudit systems}, 
      author={Doga Murat Kurkcuoglu and M. Sohaib Alam and Joshua Adam Job and Andy C. Y. Li and Alexandru Macridin and Gabriel N. Perdue and Stephen Providence},
      year={2022},
      eprint={2108.13357},
      archivePrefix={arXiv},
      primaryClass={quant-ph}
}

@article{Zohar:2012xf,
      author         = "Zohar, Erez and Cirac, J. Ignacio and Reznik, Benni",
      title          = "{Cold-Atom Quantum Simulator for SU(2) Yang-Mills Lattice
                        Gauge Theory}",
      journal        = "Phys. Rev. Lett.",
      volume         = "110",
      year           = "2013",
      number         = "12",
      pages          = "125304",
      doi            = "10.1103/PhysRevLett.110.125304",
      eprint         = "1211.2241",
      archivePrefix  = "arXiv",
      primaryClass   = "quant-ph",
      SLACcitation   = "%%CITATION = ARXIV:1211.2241;%%"
}

@article{Illa:2024kmf,
    author = "Illa, Marc and Robin, Caroline E. P. and Savage, Martin J.",
    title = "{Qu8its for quantum simulations of lattice quantum chromodynamics}",
    eprint = "2403.14537",
    archivePrefix = "arXiv",
    primaryClass = "quant-ph",
    reportNumber = "IQuS@UW-21-074",
    doi = "10.1103/PhysRevD.110.014507",
    journal = "Phys. Rev. D",
    volume = "110",
    number = "1",
    pages = "014507",
    year = "2024"
}

@article{Singh:2019jog,
    author = "Singh, Hersh",
    title = "{Qubit regularized O(N) nonlinear sigma models}",
    eprint = "1911.12353",
    archivePrefix = "arXiv",
    primaryClass = "hep-lat",
    reportNumber = "IQuS@UW-21-021,INT-PUB-22-007",
    doi = "10.1103/PhysRevD.105.114509",
    journal = "Phys. Rev. D",
    volume = "105",
    number = "11",
    pages = "114509",
    year = "2022"
}

@article{Alexandru:2023qzd,
    author = "Alexandru, Andrei and Bedaque, Paulo F. and Carosso, Andrea and Cervia, Michael J. and Murairi, Edison M. and Sheng, Andy",
    title = "{Fuzzy gauge theory for quantum computers}",
    eprint = "2308.05253",
    archivePrefix = "arXiv",
    primaryClass = "hep-lat",
    doi = "10.1103/PhysRevD.109.094502",
    journal = "Phys. Rev. D",
    volume = "109",
    number = "9",
    pages = "094502",
    year = "2024"
}

@article{ORLAND1990647,
	author = {Peter Orland and Daniel Rohrlich},
	doi = {https://doi.org/10.1016/0550-3213(90)90646-U},
	issn = {0550-3213},
	journal = {Nuclear Physics B},
	number = {3},
	pages = {647-672},
	title = {Lattice gauge magnets: Local isospin from spin},
	url = {https://www.sciencedirect.com/science/article/pii/055032139090646U},
	volume = {338},
	year = {1990}
}

@misc{Kane:2025ybw,
    author = "Kane, Christopher F. and Hariprakash, Siddharth and Bauer, Christian W.",
    title = "{Obtaining continuum physics from dynamical simulations of Hamiltonian lattice gauge theories}",
    eprint = "2506.16559",
    archivePrefix = "arXiv",
    primaryClass = "hep-lat",
    month = "6",
    year = "2025"
}

@misc{Osborne:2023rzx,
    author = "Osborne, Jesse and Yang, Bing and McCulloch, Ian P. and Hauke, Philipp and Halimeh, Jad C.",
    title = "{Spin-$S$$\mathrm{U}(1)$ Quantum Link Models with Dynamical Matter on a Quantum Simulator}",
    eprint = "2305.06368",
    archivePrefix = "arXiv",
    primaryClass = "cond-mat.quant-gas",
    month = "5",
    year = "2023"
}

@phdthesis{Kadam:2023gli,
    author = "Kadam, Saurabh Vasant",
    title = "{Theoretical Developments in Lattice Gauge Theory for Applications in Double-beta Decay Processes and Quantum Simulation}",
    eprint = "2312.00780",
    archivePrefix = "arXiv",
    primaryClass = "hep-lat",
    doi = "10.13016/dspace/cvbq-c4jt",
    school = "Maryland U., College Park",
    year = "2023"
}

@misc{Fromm:2023bit,
    author = "Fromm, Michael and Philipsen, Owe and Unger, Wolfgang and Winterowd, Christopher",
    title = "{Quantum Gate Sets for Lattice QCD in the strong coupling limit: $N_f=1$}",
    eprint = "2308.03196",
    archivePrefix = "arXiv",
    primaryClass = "hep-lat",
    month = "8",
    year = "2023"
}

@article{Davoudi:2024wyv,
    author = "Davoudi, Zohreh and Hsieh, Chung-Chun and Kadam, Saurabh V.",
    title = "{Scattering wave packets of hadrons in gauge theories: Preparation on a quantum computer}",
    eprint = "2402.00840",
    archivePrefix = "arXiv",
    primaryClass = "quant-ph",
    reportNumber = "UMD-PP-024-02, IQuS@UW-21-071",
    doi = "10.22331/q-2024-11-11-1520",
    journal = "Quantum",
    volume = "8",
    pages = "1520",
    year = "2024"
}

@article{Farrell:2024fit,
    author = "Farrell, Roland C. and Illa, Marc and Ciavarella, Anthony N. and Savage, Martin J.",
    title = "{Quantum simulations of hadron dynamics in the Schwinger model using 112 qubits}",
    eprint = "2401.08044",
    archivePrefix = "arXiv",
    primaryClass = "quant-ph",
    reportNumber = "IQuS@UW-21-069, NT@UW-24-1",
    doi = "10.1103/PhysRevD.109.114510",
    journal = "Phys. Rev. D",
    volume = "109",
    number = "11",
    pages = "114510",
    year = "2024"
}

@article{Calajo:2024qrc,
    author = "Calaj{\`o}, Giuseppe and Magnifico, Giuseppe and Edmunds, Claire and Ringbauer, Martin and Montangero, Simone and Silvi, Pietro",
    title = "{Digital Quantum Simulation of a (1+1)D SU(2) Lattice Gauge Theory with Ion Qudits}",
    eprint = "2402.07987",
    archivePrefix = "arXiv",
    primaryClass = "quant-ph",
    doi = "10.1103/PRXQuantum.5.040309",
    journal = "PRX Quantum",
    volume = "5",
    number = "4",
    pages = "040309",
    year = "2024"
}

@misc{Kan:2021xfc,
    author = "Kan, Angus and Nam, Yunseong",
    title = "{Lattice Quantum Chromodynamics and Electrodynamics on a Universal Quantum Computer}",
    eprint = "2107.12769",
    archivePrefix = "arXiv",
    primaryClass = "quant-ph",
    month = "7",
    year = "2021"
}

@article{deJong:2021wsd,
    author = "de Jong, Wibe A. and Lee, Kyle and Mulligan, James and P{\l}osko{\'n}, Mateusz and Ringer, Felix and Yao, Xiaojun",
    title = "{Quantum simulation of nonequilibrium dynamics and thermalization in the Schwinger model}",
    eprint = "2106.08394",
    archivePrefix = "arXiv",
    primaryClass = "quant-ph",
    reportNumber = "MIT-CTP/5308",
    doi = "10.1103/PhysRevD.106.054508",
    journal = "Phys. Rev. D",
    volume = "106",
    number = "5",
    pages = "054508",
    year = "2022"
}

@article{Cohen:2021imf,
    author = "Cohen, Thomas D. and Lamm, Henry and Lawrence, Scott and Yamauchi, Yukari",
    collaboration = "NuQS",
    title = "{Quantum algorithms for transport coefficients in gauge theories}",
    eprint = "2104.02024",
    archivePrefix = "arXiv",
    primaryClass = "hep-lat",
    reportNumber = "FERMILAB-PUB-21-091-T",
    doi = "10.1103/PhysRevD.104.094514",
    journal = "Phys. Rev. D",
    volume = "104",
    number = "9",
    pages = "094514",
    year = "2021"
}

@inproceedings{Zohar:2018nvl,
    author = "Emonts, Patrick and Zohar, Erez",
    title = "{Gauss law, Minimal Coupling and Fermionic PEPS for Lattice Gauge Theories}",
    booktitle = "{Tensor Network and entanglement}",
    eprint = "1807.01294",
    archivePrefix = "arXiv",
    primaryClass = "quant-ph",
    doi = "10.21468/SciPostPhysLectNotes.12",
    month = "7",
    year = "2018"
}

@article{Gustafson:2023ayr,
    author = "Gustafson, Erik J. and Lamm, Henry and Unmuth-Yockey, Judah",
    title = "{Quantum mean estimation for lattice field theory}",
    eprint = "2303.00094",
    archivePrefix = "arXiv",
    primaryClass = "hep-lat",
    reportNumber = "FERMILAB-PUB-23-095-QIS-T",
    doi = "10.1103/PhysRevD.107.114511",
    journal = "Phys. Rev. D",
    volume = "107",
    number = "11",
    pages = "114511",
    year = "2023"
}

@misc{Davoudi:2020yln,
    author = "Davoudi, Zohreh and Raychowdhury, Indrakshi and Shaw, Andrew",
    title = "{Search for Efficient Formulations for Hamiltonian Simulation of non-Abelian Lattice Gauge Theories}",
    eprint = "2009.11802",
    archivePrefix = "arXiv",
    primaryClass = "hep-lat",
    reportNumber = "UMD-PP-020-6",
    month = "9",
    year = "2020"
}

@article{Davoudi:2022uzo,
    author = "Davoudi, Zohreh and Mueller, Niklas and Powers, Connor",
    title = "{Towards Quantum Computing Phase Diagrams of Gauge Theories with Thermal Pure Quantum States}",
    eprint = "2208.13112",
    archivePrefix = "arXiv",
    primaryClass = "hep-lat",
    doi = "10.1103/PhysRevLett.131.081901",
    journal = "Phys. Rev. Lett.",
    volume = "131",
    number = "8",
    pages = "081901",
    year = "2023"
}

@article{Avkhadiev:2019niu,
    author = "Avkhadiev, A. and Shanahan, P. E. and Young, R. D.",
    title = "{Accelerating Lattice Quantum Field Theory Calculations via Interpolator Optimization Using Noisy Intermediate-Scale Quantum Computing}",
    eprint = "1908.04194",
    archivePrefix = "arXiv",
    primaryClass = "hep-lat",
    reportNumber = "MIT-CTP/5138; ADP-19-17/T1097",
    doi = "10.1103/PhysRevLett.124.080501",
    journal = "Phys. Rev. Lett.",
    volume = "124",
    number = "8",
    pages = "080501",
    year = "2020"
}

@article{Singh:2019uwd,
    author = "Singh, Hersh and Chandrasekharan, Shailesh",
    title = "{Qubit regularization of the $O(3)$ sigma model}",
    eprint = "1905.13204",
    archivePrefix = "arXiv",
    primaryClass = "hep-lat",
    doi = "10.1103/PhysRevD.100.054505",
    journal = "Phys. Rev. D",
    volume = "100",
    number = "5",
    pages = "054505",
    year = "2019"
}

@article{Gui:2023coj,
    author = "Gui, Kaiwen and Dalzell, Alexander M. and Achille, Alessandro and Suchara, Martin and Chong, Frederic T.",
    title = "{Spacetime-Efficient Low-Depth Quantum State Preparation with Applications}",
    eprint = "2303.02131",
    archivePrefix = "arXiv",
    primaryClass = "quant-ph",
    doi = "10.22331/q-2024-02-15-1257",
    journal = "Quantum",
    volume = "8",
    pages = "1257",
    year = "2024"
}

@article{Clancy:2023kla,
    author = "Clancy, Michael",
    title = "{Toward a Ginsparg-Wilson lattice Hamiltonian}",
    eprint = "2312.08647",
    archivePrefix = "arXiv",
    primaryClass = "hep-lat",
    reportNumber = "INT-PUB-23-051",
    doi = "10.1103/PhysRevD.110.L011502",
    journal = "Phys. Rev. D",
    volume = "110",
    number = "1",
    pages = "L011502",
    year = "2024"
}

@article{Florio:2023kel,
    author = "Florio, Adrien and Weichselbaum, Andreas and Valgushev, Semeon and Pisarski, Robert D.",
    title = "{Mass gaps of a Z3 gauge theory with three fermion flavors in 1+1 dimensions}",
    eprint = "2310.18312",
    archivePrefix = "arXiv",
    primaryClass = "hep-th",
    doi = "10.1103/PhysRevD.110.045013",
    journal = "Phys. Rev. D",
    volume = "110",
    number = "4",
    pages = "045013",
    year = "2024"
}

@article{PhysRevD.107.114513,
  title = {Hamiltonians and gauge-invariant Hilbert space for lattice Yang-Mills-like theories with finite gauge group},
  author = {Mariani, A. and Pradhan, S. and Ercolessi, E.},
  journal = {Phys. Rev. D},
  volume = {107},
  issue = {11},
  pages = {114513},
  numpages = {15},
  year = {2023},
  month = {Jun},
  publisher = {American Physical Society},
  doi = {10.1103/PhysRevD.107.114513},
  url = {https://link.aps.org/doi/10.1103/PhysRevD.107.114513}
}

@article{Ciavarella:2024lsp,
    author = "Ciavarella, Anthony N.",
    title = "{String breaking in the heavy quark limit with scalable circuits}",
    eprint = "2411.05915",
    archivePrefix = "arXiv",
    primaryClass = "quant-ph",
    doi = "10.1103/PhysRevD.111.054501",
    journal = "Phys. Rev. D",
    volume = "111",
    number = "5",
    pages = "054501",
    year = "2025"
}

@article{Ciavarella:2021nmj,
    author = "Ciavarella, Anthony and Klco, Natalie and Savage, Martin J.",
    title = "{Trailhead for quantum simulation of SU(3) Yang-Mills lattice gauge theory in the local multiplet basis}",
    eprint = "2101.10227",
    archivePrefix = "arXiv",
    primaryClass = "quant-ph",
    reportNumber = "IQuS@UW-21-001",
    doi = "10.1103/PhysRevD.103.094501",
    journal = "Phys. Rev. D",
    volume = "103",
    number = "9",
    pages = "094501",
    year = "2021"
}

@misc{Gidney:2024alh,
    author = "Gidney, Craig and Shutty, Noah and Jones, Cody",
    title = "{Magic state cultivation: growing T states as cheap as CNOT gates}",
    eprint = "2409.17595",
    archivePrefix = "arXiv",
    primaryClass = "quant-ph",
    month = "9",
    year = "2024"
}

@article{Rhodes:2024zbr,
    author = "Rhodes, Mason L. and Kreshchuk, Michael and Pathak, Shivesh",
    title = "{Exponential Improvements in the Simulation of Lattice Gauge Theories Using Near-Optimal Techniques}",
    eprint = "2405.10416",
    archivePrefix = "arXiv",
    primaryClass = "quant-ph",
    doi = "10.1103/PRXQuantum.5.040347",
    journal = "PRX Quantum",
    volume = "5",
    number = "4",
    pages = "040347",
    year = "2024"
}

@article{Grimus:2010ak,
    author = "Grimus, W. and Ludl, P. O.",
    title = "{Principal series of finite subgroups of SU(3)}",
    eprint = "1006.0098",
    archivePrefix = "arXiv",
    primaryClass = "hep-ph",
    reportNumber = "UWTHPH-2010-10",
    doi = "10.1088/1751-8113/43/44/445209",
    journal = "J. Phys. A",
    volume = "43",
    pages = "445209",
    year = "2010"
}

@misc{Di:2011cvl,
    author = "Di, Yao-Min and Wei, Hai-Rui",
    title = "{Elementary gates for ternary quantum logic circuit}",
    eprint = "1105.5485",
    archivePrefix = "arXiv",
    primaryClass = "quant-ph",
    month = "5",
    year = "2011"
}

@article{Alexandru:2019nsa,
    author = "Alexandru, Andrei and Bedaque, Paulo F. and Harmalkar, Siddhartha and Lamm, Henry and Lawrence, Scott and Warrington, Neill C.",
    archivePrefix = "arXiv",
    collaboration = "NuQS",
    doi = "10.1103/PhysRevD.100.114501",
    eprint = "1906.11213",
    journal = "Phys.Rev.D",
    number = "11",
    pages = "114501",
    primaryClass = "hep-lat",
    title = "Gluon Field Digitization for Quantum Computers",
    volume = "100",
    year = "2019"
}

@article{Bhanot:1981pj,
      author         = "Bhanot, Gyan",
      title          = "{SU(3) Lattice Gauge Theory in Four-dimensions With a
                        Modified Wilson Action}",
      journal        = "Phys. Lett.",
      volume         = "108B",
      year           = "1982",
      pages          = "337-340",
      doi            = "10.1016/0370-2693(82)91207-2",
      reportNumber   = "Print-81-0779 (IAS,PRINCETON), BNL-30161",
      SLACcitation   = "%%CITATION = PHLTA,108B,337;%%"
}

@article{Creutz:1982dn,
      author         = "Creutz, Michael and Okawa, Masanori",
      title          = "{Generalized Actions in $Z(p$) Lattice Gauge Theory}",
      journal        = "Nucl. Phys.",
      volume         = "B220",
      year           = "1983",
      pages          = "149-166",
      doi            = "10.1016/0550-3213(83)90220-1",
      reportNumber   = "BNL-32052",
      SLACcitation   = "%%CITATION = NUPHA,B220,149;%%"
}

@article{Armon:2021uqr,
    author = "Armon, Tsafrir and Ashkenazi, Shachar and Garc{\'\i}a-Moreno, Gerardo and Gonz{\'a}lez-Tudela, Alejandro and Zohar, Erez",
    title = "{Photon-Mediated Stroboscopic Quantum Simulation of a Z2 Lattice Gauge Theory}",
    eprint = "2107.13024",
    archivePrefix = "arXiv",
    primaryClass = "quant-ph",
    doi = "10.1103/PhysRevLett.127.250501",
    journal = "Phys. Rev. Lett.",
    volume = "127",
    number = "25",
    pages = "250501",
    year = "2021"
}

@article{Lamm:2019bik,
      author         = "Lamm, Henry and Lawrence, Scott and Yamauchi, Yukari",
      title          = "{General Methods for Digital Quantum Simulation of Gauge
                        Theories}",
      collaboration  = "NuQS",
      journal        = "Phys. Rev.",
      volume         = "D100",
      year           = "2019",
      number         = "3",
      pages          = "034518",
      doi            = "10.1103/PhysRevD.100.034518",
      eprint         = "1903.08807",
      archivePrefix  = "arXiv",
      primaryClass   = "hep-lat",
      SLACcitation   = "%%CITATION = ARXIV:1903.08807;%%"
}

@article{Luo:2019vmi,
    author = "Luo, Di and Shen, Jiayu and Highman, Michael and Clark, Bryan K. and DeMarco, Brian and El-Khadra, Aida X. and Gadway, Bryce",
    title = "{Framework for simulating gauge theories with dipolar spin systems}",
    eprint = "1912.11488",
    archivePrefix = "arXiv",
    primaryClass = "quant-ph",
    doi = "10.1103/PhysRevA.102.032617",
    journal = "Phys. Rev. A",
    volume = "102",
    number = "3",
    pages = "032617",
    year = "2020"
}

@article{Brower:2020huh,
    author = "Brower, Richard C. and Berenstein, David and Kawai, Hiroki",
    title = "{Lattice Gauge Theory for a Quantum Computer}",
    eprint = "2002.10028",
    archivePrefix = "arXiv",
    primaryClass = "hep-lat",
    journal = "PoS",
    volume = "LATTICE2019",
    pages = "112",
    year = "2019"
}

@misc{Liu:2020eoa,
    author = "Liu, Junyu and Xin, Yuan",
    title = "{Quantum simulation of quantum field theories as quantum chemistry}",
    eprint = "2004.13234",
    archivePrefix = "arXiv",
    primaryClass = "hep-th",
    reportNumber = "CALT-TH-2020-009",
    month = "4",
    year = "2020"
}

@misc{Kreshchuk:2020kcz,
    author = "Kreshchuk, Michael and Jia, Shaoyang and Kirby, William M. and Goldstein, Gary and Vary, James P. and Love, Peter J.",
    title = "{Light-Front Field Theory on Current Quantum Computers}",
    eprint = "2009.07885",
    archivePrefix = "arXiv",
    primaryClass = "quant-ph",
    month = "9",
    year = "2020"
}

@misc{Kreshchuk:2020dla,
    author = "Kreshchuk, Michael and Kirby, William M. and Goldstein, Gary and Beauchemin, Hugo and Love, Peter J.",
    title = "{Quantum Simulation of Quantum Field Theory in the Light-Front Formulation}",
    eprint = "2002.04016",
    archivePrefix = "arXiv",
    primaryClass = "quant-ph",
    month = "2",
    year = "2020"
}

@article{Haase:2020kaj,
    author = "Haase, Jan F. and Dellantonio, Luca and Celi, Alessio and Paulson, Danny and Kan, Angus and Jansen, Karl and Muschik, Christine A.",
    title = "{A resource efficient approach for quantum and classical simulations of gauge theories in particle physics}",
    eprint = "2006.14160",
    archivePrefix = "arXiv",
    primaryClass = "quant-ph",
    reportNumber = "DESY-20-146",
    doi = "10.22331/q-2021-02-04-393",
    journal = "Quantum",
    volume = "5",
    pages = "393",
    year = "2021"
}

@article{Zhang:2018ufj,
      author         = "Zhang, Jin and Unmuth-Yockey, J. and Zeiher, J. and
                        Bazavov, A. and Tsai, S. -W. and Meurice, Y.",
      title          = "{Quantum simulation of the universal features of the
                        Polyakov loop}",
      journal        = "Phys. Rev. Lett.",
      volume         = "121",
      year           = "2018",
      number         = "22",
      pages          = "223201",
      doi            = "10.1103/PhysRevLett.121.223201",
      eprint         = "1803.11166",
      archivePrefix  = "arXiv",
      primaryClass   = "hep-lat",
      SLACcitation   = "%%CITATION = ARXIV:1803.11166;%%"
}

@article{Bazavov:2015kka,
      author         = "Bazavov, Alexei and Meurice, Yannick and Tsai, Shan-Wen
                        and Unmuth-Yockey, Judah and Zhang, Jin",
      title          = "{Gauge-invariant implementation of the Abelian Higgs
                        model on optical lattices}",
      journal        = "Phys. Rev.",
      volume         = "D92",
      year           = "2015",
      number         = "7",
      pages          = "076003",
      doi            = "10.1103/PhysRevD.92.076003",
      eprint         = "1503.08354",
      archivePrefix  = "arXiv",
      primaryClass   = "hep-lat",
      reportNumber   = "INT-PUB-15-008",
      SLACcitation   = "%%CITATION = ARXIV:1503.08354;%%"
}

@article{Unmuth-Yockey:2018xak,
    author = "Unmuth-Yockey, Judah F.",
    archivePrefix = "arXiv",
    doi = "10.1103/PhysRevD.99.074502",
    eprint = "1811.05884",
    journal = "Phys.\ Rev.\ D",
    number = "7",
    pages = "074502",
    primaryClass = "hep-lat",
    title = "{Gauge-invariant rotor Hamiltonian from dual variables of 3D $U(1)$ gauge theory}",
    volume = "99",
    year = "2019"
}

@article{Unmuth-Yockey:2018ugm,
      author         = "Unmuth-Yockey, Judah and Zhang, Jin and Bazavov, Alexei
                        and Meurice, Yannick and Tsai, Shan-Wen",
      title          = "{Universal features of the Abelian Polyakov loop in 1+1
                        dimensions}",
      journal        = "Phys. Rev.",
      volume         = "D98",
      year           = "2018",
      number         = "9",
      pages          = "094511",
      doi            = "10.1103/PhysRevD.98.094511",
      eprint         = "1807.09186",
      archivePrefix  = "arXiv",
      primaryClass   = "hep-lat",
      SLACcitation   = "%%CITATION = ARXIV:1807.09186;%%"
}

@article{Gustafson:2019mpk,
    author = "Gustafson, Erik and Meurice, Yannick and Unmuth-Yockey, Judah",
    title = "{Quantum simulation of scattering in the quantum Ising model}",
    eprint = "1901.05944",
    archivePrefix = "arXiv",
    primaryClass = "hep-lat",
    doi = "10.1103/PhysRevD.99.094503",
    journal = "Phys. Rev. D",
    volume = "99",
    number = "9",
    pages = "094503",
    year = "2019"
}

@article{Gustafson:2021qbt,
    author = "Gustafson, Erik",
    title = "{Prospects for Simulating a Qudit Based Model of (1+1)d Scalar QED}",
    eprint = "2104.10136",
    archivePrefix = "arXiv",
    primaryClass = "quant-ph",
    doi = "10.1103/PhysRevD.103.114505",
    journal = "Phys. Rev. D",
    volume = "103",
    number = "11",
    pages = "114505",
    year = "2021"
}

@misc{Kreshchuk:2020aiq,
    author = "Kreshchuk, Michael and Jia, Shaoyang and Kirby, William M. and Goldstein, Gary and Vary, James P. and Love, Peter J.",
    title = "{Simulating Hadronic Physics on NISQ devices using Basis Light-Front Quantization}",
    eprint = "2011.13443",
    archivePrefix = "arXiv",
    primaryClass = "quant-ph",
    month = "11",
    year = "2020"
}

@article{Klco:2019evd,
    author = "Klco, Natalie and Stryker, Jesse R. and Savage, Martin J.",
    title = "{SU(2) non-Abelian gauge field theory in one dimension on digital quantum computers}",
    eprint = "1908.06935",
    archivePrefix = "arXiv",
    primaryClass = "quant-ph",
    reportNumber = "INT-PUB-19-033",
    doi = "10.1103/PhysRevD.101.074512",
    journal = "Phys. Rev. D",
    volume = "101",
    number = "7",
    pages = "074512",
    year = "2020"
}

@article{Gustafson:2020yfe,
    author = "Gustafson, Erik J. and Lamm, Henry",
    title = "{Toward quantum simulations of $\mathbb{Z}_2$ gauge theory without state preparation}",
    eprint = "2011.11677",
    archivePrefix = "arXiv",
    primaryClass = "hep-lat",
    reportNumber = "FERMILAB-PUB-20-611-T",
    doi = "10.1103/PhysRevD.103.054507",
    journal = "Phys. Rev. D",
    volume = "103",
    number = "5",
    pages = "054507",
    year = "2021"
}

@article{Gustafson:2019vsd,
    author = "Gustafson, Erik and Dreher, Patrick and Hang, Zheyue and Meurice, Yannick",
    title = "{Benchmarking quantum computers for real-time evolution of a $(1+1)$ field theory with error mitigation}",
    eprint = "1910.09478",
    archivePrefix = "arXiv",
    primaryClass = "hep-lat",
    journal = "Quantum Sci. Technol.",
    volume = "6",
    pages = "045020",
    year = "2021"
}

@misc{Spagnoli:2024mib,
    author = "Spagnoli, L. and Roggero, A. and Wiebe, N.",
    title = "{Fault-tolerant simulation of Lattice Gauge Theories with gauge covariant codes}",
    eprint = "2405.19293",
    archivePrefix = "arXiv",
    primaryClass = "quant-ph",
    month = "5",
    year = "2024"
}

@misc{Kurkcuoglu:2025gik,
    author = {K{\"u}rk{\c{c}}{\"u}oglu, Doga Murat and Lamm, Henry and Ogunkoya, Oluwadara and Pierattelli, Leonardo},
    title = "{The 2T-quoctit: a two-mode bosonic qudit for high energy physics}",
    eprint = "2509.21941",
    archivePrefix = "arXiv",
    primaryClass = "quant-ph",
    reportNumber = "FERMILAB-PUB-25-0143-SQMS-STUDENT-T",
    month = "9",
    year = "2025"
}

@misc{Haruna:2025piy,
    author = "Haruna, Junichi",
    title = "{Note on Logical Gates by Gauge Field Formalism of Quantum Error Correction}",
    eprint = "2511.15224",
    archivePrefix = "arXiv",
    primaryClass = "hep-th",
    month = "11",
    year = "2025"
}

@article{PhysRevLett.114.080502,
  title = {Efficient Synthesis of Universal Repeat-Until-Success Quantum Circuits},
  author = {Bocharov, Alex and Roetteler, Martin and Svore, Krysta M.},
  journal = {Phys. Rev. Lett.},
  volume = {114},
  issue = {8},
  pages = {080502},
  numpages = {5},
  year = {2015},
  month = {Feb},
  publisher = {American Physical Society},
  doi = {10.1103/PhysRevLett.114.080502},
  url = {https://link.aps.org/doi/10.1103/PhysRevLett.114.080502}
}

@article{peruzzo2014variational,
  title={A variational eigenvalue solver on a photonic quantum processor},
  author={Peruzzo, Alberto and McClean, Jarrod and Shadbolt, Peter and Yung, Man-Hong and Zhou, Xiao-Qi and Love, Peter J and Aspuru-Guzik, Al{\'a}n and O’brien, Jeremy L},
  journal={Nature communications},
  volume={5},
  pages={4213},
  year={2014},
  publisher={Nature Publishing Group}
}

@article{Zache:2023cfj,
    author = "Zache, Torsten V. and Gonz\'alez-Cuadra, Daniel and Zoller, Peter",
    title = "{Fermion-qudit quantum processors for simulating lattice gauge theories with matter}",
    eprint = "2303.08683",
    archivePrefix = "arXiv",
    primaryClass = "quant-ph",
    doi = "10.22331/q-2023-10-16-1140",
    journal = "Quantum",
    volume = "7",
    pages = "1140",
    year = "2023"
}

@article{Zache:2023dko,
    author = "Zache, Torsten V. and Gonz\'alez-Cuadra, Daniel and Zoller, Peter",
    title = "{Quantum and Classical Spin-Network Algorithms for q-Deformed Kogut-Susskind Gauge Theories}",
    eprint = "2304.02527",
    archivePrefix = "arXiv",
    primaryClass = "quant-ph",
    doi = "10.1103/PhysRevLett.131.171902",
    journal = "Phys. Rev. Lett.",
    volume = "131",
    number = "17",
    pages = "171902",
    year = "2023"
}

@article{Bender:2018rdp,
      author         = "Bender, Julian and Zohar, Erez and Farace, Alessandro and
                        Cirac, J. Ignacio",
      title          = "{Digital quantum simulation of lattice gauge theories in
                        three spatial dimensions}",
      journal        = "New J. Phys.",
      volume         = "20",
      year           = "2018",
      number         = "9",
      pages          = "093001",
      doi            = "10.1088/1367-2630/aadb71",
      eprint         = "1804.02082",
      archivePrefix  = "arXiv",
      primaryClass   = "quant-ph",
      SLACcitation   = "%%CITATION = ARXIV:1804.02082;%%"
}

@misc{kitaev,
    author = "Kitaev, A. Yu.",
    title = "{Quantum measurements and the Abelian stabilizer problem}",
    eprint = "quant-ph/9511026",
    archivePrefix = "arXiv",
    month = "11",
    year = "1995",
    url={https://arxiv.org/abs/quant-ph/9511026}
}

@book{NC,
  title={Quantum computation and quantum information},
  author={Nielsen, Michael A and Chuang, Isaac},
  year={2002},
  publisher={Cambridge University Press}
}

@inproceedings{hales2000improved,
  title={An improved quantum Fourier transform algorithm and applications},
  author={Hales, Lisa and Hallgren, Sean},
  booktitle={Proceedings 41st Annual Symposium on Foundations of Computer Science},
  pages={515--525},
  year={2000},
  organization={IEEE}
}

@article{Gustafson:2023kvd,
    author = "Gustafson, Erik J. and Lamm, Henry and Lovelace, Felicity",
    title = "{Primitive quantum gates for an SU(2) discrete subgroup: Binary octahedral}",
    eprint = "2312.10285",
    archivePrefix = "arXiv",
    primaryClass = "hep-lat",
    reportNumber = "FERMILAB-PUB-23-753-SQMS-T",
    doi = "10.1103/PhysRevD.109.054503",
    journal = "Phys. Rev. D",
    volume = "109",
    number = "5",
    pages = "054503",
    year = "2024"
}

@article{Zohar:2013zla,
      author         = "Zohar, Erez and Cirac, J. Ignacio and Reznik, Benni",
      title          = "{Quantum simulations of gauge theories with ultracold
                        atoms: local gauge invariance from angular momentum
                        conservation}",
      journal        = "Phys. Rev.",
      volume         = "A88",
      year           = "2013",
      pages          = "023617",
      doi            = "10.1103/PhysRevA.88.023617",
      eprint         = "1303.5040",
      archivePrefix  = "arXiv",
      primaryClass   = "quant-ph",
      SLACcitation   = "%%CITATION = ARXIV:1303.5040;%%"
}

@article{Brower:1997ha,
	Archiveprefix = {arXiv},
	Author = {Brower, R. and Chandrasekharan, S. and Wiese, U. J.},
	Doi = {10.1103/PhysRevD.60.094502},
	Eprint = {hep-th/9704106},
	Journal = {Phys. Rev.},
	Pages = {094502},
	Primaryclass = {hep-th},
	Reportnumber = {MIT-CTP-2623},
	Slaccitation = {%%CITATION = HEP-TH/9704106;%%},
	Title = {{QCD as a quantum link model}},
	Volume = {D60},
	Year = {1999}}

@inproceedings{Pueschel:1998zzo,
  title={{Fast quantum Fourier transforms for a class of non-abelian groups}},
  author={P{\"u}schel, Markus and R{\"o}tteler, Martin and Beth, Thomas},
  booktitle={Applied Algebra, Algebraic Algorithms and Error-Correcting Codes: 13th International Symposium, AAECC-13 Honolulu, Hawaii, USA, November 15--19, 1999 Proceedings 13},
  pages={148--159},
  year={1999},
  organization={Springer},
  url={https://arxiv.org/abs/quant-ph/9807064}
}

@article{moore2003generic,
  title={{Generic quantum Fourier transforms}},
  author={Moore, Cristopher and Rockmore, Daniel and Russell, Alexander},
  journal={ACM Transactions on Algorithms (TALG)},
  volume={2},
  number={4},
  pages={707--723},
  year={2006},
  publisher={ACM New York, NY, USA},
  url={https://arxiv.org/abs/quant-ph/0304064}
}

@misc{Hoyer:1997qc,
    author = "Hoyer, Peter",
    title = "{Efficient quantum transforms}",
    eprint = "quant-ph/9702028",
    archivePrefix = "arXiv",
    month = "2",
    year = "1997"
}

@inproceedings{beals1997quantum,
  title={Quantum computation of Fourier transforms over symmetric groups},
  author={Beals, Robert},
  booktitle={Proceedings of the twenty-ninth annual ACM symposium on Theory of computing},
  pages={48--53},
  year={1997}
}

@article{Carena:2022hpz,
    author = "Carena, Marcela and Gustafson, Erik J. and Lamm, Henry and Li, Ying-Ying and Liu, Wanqiang",
    title = "{Gauge theory couplings on anisotropic lattices}",
    eprint = "2208.10417",
    archivePrefix = "arXiv",
    primaryClass = "hep-lat",
    reportNumber = "FERMILAB-PUB-21-702-T",
    doi = "10.1103/PhysRevD.106.114504",
    journal = "Phys. Rev. D",
    volume = "106",
    number = "11",
    pages = "114504",
    year = "2022"
}

@article{PhysRevA.52.3457,
  title = {Elementary gates for quantum computation},
  author = {Barenco, Adriano and Bennett, Charles H. and Cleve, Richard and DiVincenzo, David P. and Margolus, Norman and Shor, Peter and Sleator, Tycho and Smolin, John A. and Weinfurter, Harald},
  journal = {Phys. Rev. A},
  volume = {52},
  issue = {5},
  pages = {3457--3467},
  numpages = {0},
  year = {1995},
  month = {Nov},
  publisher = {American Physical Society},
  doi = {10.1103/PhysRevA.52.3457},
  url = {https://link.aps.org/doi/10.1103/PhysRevA.52.3457}
}

@misc{2019arXiv190401671B,
       author = {{Baker}, Jonathan M. and {Duckering}, Casey and {Hoover}, Alexander and {Chong}, Frederic T.},
        title = "{Decomposing Quantum Generalized Toffoli with an Arbitrary Number of Ancilla}",
         year = 2019,
        month = apr,
archivePrefix = "arXiv",
       eprint = "1904.01671",
 primaryClass = "quant-ph",
}

@misc{Gustafson:2023swx,
    author = "Gustafson, Erik J. and Lamm, Henry",
    title = "{Robustness of Gauge Digitization to Quantum Noise}",
    eprint = "2301.10207",
    archivePrefix = "arXiv",
    primaryClass = "hep-lat",
    reportNumber = "FERMILAB-PUB-23-018-SQMS-T",
    month = "1",
    year = "2023"
}

@article{Gattringer:2016kco,
    author = "Gattringer, Christof and Langfeld, Kurt",
    title = "{Approaches to the sign problem in lattice field theory}",
    eprint = "1603.09517",
    archivePrefix = "arXiv",
    primaryClass = "hep-lat",
    doi = "10.1142/S0217751X16430077",
    journal = "Int. J. Mod. Phys. A",
    volume = "31",
    number = "22",
    pages = "1643007",
    year = "2016"
}

@article{Clemente:2022cka,
    author = "Clemente, Giuseppe and Crippa, Arianna and Jansen, Karl",
    title = "{Strategies for the determination of the running coupling of (2+1)-dimensional QED with quantum computing}",
    eprint = "2206.12454",
    archivePrefix = "arXiv",
    primaryClass = "hep-lat",
    doi = "10.1103/PhysRevD.106.114511",
    journal = "Phys. Rev. D",
    volume = "106",
    number = "11",
    pages = "114511",
    year = "2022"
}

@article{Chuang:1996hw,
    author = "Chuang, Isaac L. and Nielsen, M. A.",
    title = "{Prescription for experimental determination of the dynamics of a quantum black box}",
    eprint = "quant-ph/9610001",
    archivePrefix = "arXiv",
    doi = "10.1080/09500349708231894",
    journal = "J. Mod. Opt.",
    volume = "44",
    pages = "2455",
    year = "1997"
}

@article{Troyer:2004ge,
      author         = "Troyer, Matthias and Wiese, Uwe-Jens",
      title          = "{Computational complexity and fundamental limitations to
                        fermionic quantum Monte Carlo simulations}",
      journal        = "Phys. Rev. Lett.",
      volume         = "94",
      year           = "2005",
      pages          = "170201",
      doi            = "10.1103/PhysRevLett.94.170201",
      eprint         = "cond-mat/0408370",
      archivePrefix  = "arXiv",
      primaryClass   = "cond-mat",
      SLACcitation   = "%%CITATION = COND-MAT/0408370;%%"
}

@article{PhysRevD.99.114507,
  title = {Tensor renormalization group study of the non-Abelian Higgs model in two dimensions},
  author = {Bazavov, Alexei and Catterall, Simon and Jha, Raghav G. and Unmuth-Yockey, Judah},
  journal = {Phys. Rev. D},
  volume = {99},
  issue = {11},
  pages = {114507},
  numpages = {12},
  year = {2019},
  month = {Jun},
  publisher = {American Physical Society},
  doi = {10.1103/PhysRevD.99.114507},
  url = {https://link.aps.org/doi/10.1103/PhysRevD.99.114507}
}

@article{Buser:2020uzs,
    author = "Buser, Alexander J. and Bhattacharya, Tanmoy and Cincio, Lukasz and Gupta, Rajan",
    title = "{State preparation and measurement in a quantum simulation of the $O$(3) sigma model}",
    eprint = "2006.15746",
    archivePrefix = "arXiv",
    primaryClass = "quant-ph",
    reportNumber = "LA-UR-20-24538",
    doi = "10.1103/PhysRevD.102.114514",
    journal = "Phys. Rev. D",
    volume = "102",
    number = "11",
    pages = "114514",
    year = "2020"
}

@article{Bhattacharya:2020gpm,
    author = "Bhattacharya, Tanmoy and Buser, Alexander J. and Chandrasekharan, Shailesh and Gupta, Rajan and Singh, Hersh",
    title = "{Qubit regularization of asymptotic freedom}",
    eprint = "2012.02153",
    archivePrefix = "arXiv",
    primaryClass = "hep-lat",
    reportNumber = "LA-UR-20-29558",
    doi = "10.1103/PhysRevLett.126.172001",
    journal = "Phys. Rev. Lett.",
    volume = "126",
    number = "17",
    pages = "172001",
    year = "2021"
}

@article{Zohar:2012ay,
      author         = "Zohar, Erez and Cirac, J. Ignacio and Reznik, Benni",
      title          = "{Simulating Compact Quantum Electrodynamics with
                        ultracold atoms: Probing confinement and nonperturbative
                        effects}",
      journal        = "Phys. Rev. Lett.",
      volume         = "109",
      year           = "2012",
      pages          = "125302",
      doi            = "10.1103/PhysRevLett.109.125302",
      eprint         = "1204.6574",
      archivePrefix  = "arXiv",
      primaryClass   = "quant-ph",
      SLACcitation   = "%%CITATION = ARXIV:1204.6574;%%"
}

@article{Wiese:2014rla,
      author         = "Wiese, Uwe-Jens",
      title          = "{Towards Quantum Simulating QCD}",
      booktitle      = "{Proceedings, 24th International Conference on
                        Ultra-Relativistic Nucleus-Nucleus Collisions (Quark
                        Matter 2014): Darmstadt, Germany, May 19-24, 2014}",
      journal        = "Nucl. Phys.",
      volume         = "A931",
      year           = "2014",
      pages          = "246-256",
      doi            = "10.1016/j.nuclphysa.2014.09.102",
      eprint         = "1409.7414",
      archivePrefix  = "arXiv",
      primaryClass   = "hep-th",
      SLACcitation   = "%%CITATION = ARXIV:1409.7414;%%"
}

@article{Zohar:2016iic,
      author         = "Zohar, Erez and Farace, Alessandro and Reznik, Benni and
                        Cirac, J. Ignacio",
      title          = "{Digital lattice gauge theories}",
      journal        = "Phys. Rev.",
      volume         = "A95",
      year           = "2017",
      number         = "2",
      pages          = "023604",
      doi            = "10.1103/PhysRevA.95.023604",
      eprint         = "1607.08121",
      archivePrefix  = "arXiv",
      primaryClass   = "quant-ph",
      SLACcitation   = "%%CITATION = ARXIV:1607.08121;%%"
}

@article{Jordan:2017lea,
    author = "Jordan, Stephen P. and Krovi, Hari and Lee, Keith S.M. and Preskill, John",
    title = "{BQP-completeness of Scattering in Scalar Quantum Field Theory}",
    eprint = "1703.00454",
    archivePrefix = "arXiv",
    primaryClass = "quant-ph",
    doi = "10.22331/q-2018-01-08-44",
    journal = "Quantum",
    volume = "2",
    pages = "44",
    year = "2018"
}

@article{Zohar:2015hwa,
      author         = "Zohar, Erez and Cirac, J. Ignacio and Reznik, Benni",
      title          = "{Quantum Simulations of Lattice Gauge Theories using
                        Ultracold Atoms in Optical Lattices}",
      journal        = "Rept. Prog. Phys.",
      volume         = "79",
      year           = "2016",
      number         = "1",
      pages          = "014401",
      doi            = "10.1088/0034-4885/79/1/014401",
      eprint         = "1503.02312",
      archivePrefix  = "arXiv",
      primaryClass   = "quant-ph",
      SLACcitation   = "%%CITATION = ARXIV:1503.02312;%%"
}

@article{Kane:2023jdo,
    author = "Kane, Christopher F. and Gomes, Niladri and Kreshchuk, Michael",
    title = "{Nearly optimal state preparation for quantum simulations of lattice gauge theories}",
    eprint = "2310.13757",
    archivePrefix = "arXiv",
    primaryClass = "quant-ph",
    reportNumber = "Phys.Rev.A 110 (2024) 1, 012455",
    doi = "10.1103/PhysRevA.110.012455",
    journal = "Phys. Rev. A",
    volume = "110",
    number = "1",
    pages = "012455",
    year = "2024"
}

@article{Gustafson:2022xdt,
    author = "Gustafson, Erik J. and Lamm, Henry and Lovelace, Felicity and Musk, Damian",
    title = "{Primitive quantum gates for an SU(2) discrete subgroup: Binary tetrahedral}",
    eprint = "2208.12309",
    archivePrefix = "arXiv",
    primaryClass = "quant-ph",
    reportNumber = "FERMILAB-PUB-22-583-SQMS-T",
    doi = "10.1103/PhysRevD.106.114501",
    journal = "Phys. Rev. D",
    volume = "106",
    number = "11",
    pages = "114501",
    year = "2022"
}

@misc{Gustafson:2022hjf,
    author = "Gustafson, Erik J.",
    title = "{Stout Smearing on a Quantum Computer}",
    eprint = "2211.05607",
    archivePrefix = "arXiv",
    primaryClass = "hep-lat",
    reportNumber = "FERMILAB-PUB-22-838-T",
    month = "11",
    year = "2022"
}

@article{Xie:2022jgj,
	archiveprefix = {arXiv},
	author = {Xie, Xu-Dan and Guo, Xingyu and Xing, Hongxi and Xue, Zheng-Yuan and Zhang, Dan-Bo and Zhu, Shi-Liang},
	collaboration = {QuNu},
	doi = {10.1103/PhysRevD.106.054509},
	eprint = {2205.12767},
	journal = {Phys. Rev. D},
	number = {5},
	pages = {054509},
	primaryclass = {quant-ph},
	title = {{Variational thermal quantum simulation of the lattice Schwinger model}},
	volume = {106},
	year = {2022}
}

@article{motta2020determining,
  title={Determining eigenstates and thermal states on a quantum computer using quantum imaginary time evolution},
  author={Motta, Mario and Sun, Chong and Tan, Adrian TK and O’Rourke, Matthew J and Ye, Erika and Minnich, Austin J and Brandao, Fernando GSL and Chan, Garnet Kin-Lic},
  journal={Nature Physics},
  volume={16},
  number={2},
  pages={205--210},
  year={2020},
  publisher={Nature Publishing Group}
}

@article{PhysRevLett.108.080402,
  title = {Thermalization in Nature and on a Quantum Computer},
  author = {Riera, Arnau and Gogolin, Christian and Eisert, Jens},
  journal = {Phys. Rev. Lett.},
  volume = {108},
  issue = {8},
  pages = {080402},
  numpages = {5},
  year = {2012},
  month = {Feb},
  publisher = {American Physical Society},
  doi = {10.1103/PhysRevLett.108.080402},
  url = {https://link.aps.org/doi/10.1103/PhysRevLett.108.080402}
}

@article{Charles:2023zbl,
    author = "Charles, Clement and Gustafson, Erik J. and Hardt, Elizabeth and Herren, Florian and Hogan, Norman and Lamm, Henry and Starecheski, Sara and Van de Water, Ruth S. and Wagman, Michael L.",
    title = "{Simulating Z2 lattice gauge theory on a quantum computer}",
    eprint = "2305.02361",
    archivePrefix = "arXiv",
    primaryClass = "hep-lat",
    reportNumber = "FERMILAB-PUB-23-171-SQMS-T",
    doi = "10.1103/PhysRevE.109.015307",
    journal = "Phys. Rev. E",
    volume = "109",
    number = "1",
    pages = "015307",
    year = "2024"
}

@article{Clemente:2020lpr,
    author = "Clemente, Giuseppe and others",
    collaboration = "QuBiPF",
    title = "{Quantum computation of thermal averages in the presence of a sign problem}",
    eprint = "2001.05328",
    archivePrefix = "arXiv",
    primaryClass = "hep-lat",
    doi = "10.1103/PhysRevD.101.074510",
    journal = "Phys. Rev. D",
    volume = "101",
    number = "7",
    pages = "074510",
    year = "2020"
}

@article{PhysRevD.11.395,
  title = {Hamiltonian formulation of {W}ilson's lattice gauge theories},
  author = {Kogut, John and Susskind, Leonard},
  journal = {Phys. Rev. D},
  volume = {11},
  issue = {2},
  pages = {395--408},
  numpages = {0},
  year = {1975},
  month = {Jan},
  publisher = {American Physical Society},
  doi = {10.1103/PhysRevD.11.395},
  url = {https://link.aps.org/doi/10.1103/PhysRevD.11.395}
}

@misc{Crippa:2024cqr,
    author = {Crippa, Arianna and Romiti, Simone and Funcke, Lena and Jansen, Karl and K\"uhn, Stefan and Stornati, Paolo and Urbach, Carsten},
    title = "{Towards determining the (2+1)-dimensional Quantum Electrodynamics running coupling with Monte Carlo and quantum computing methods}",
    eprint = "2404.17545",
    archivePrefix = "arXiv",
    primaryClass = "hep-lat",
    month = "4",
    year = "2024"
}

@article{Funcke:2022opx,
    author = {Funcke, Lena and Gro\ss{}, Christiane Franziska and Jansen, Karl and K\"uhn, Stefan and Romiti, Simone and Urbach, Carsten},
    title = "{Hamiltonian limit of lattice QED in 2+1 dimensions}",
    eprint = "2212.09627",
    archivePrefix = "arXiv",
    primaryClass = "hep-lat",
    reportNumber = "MIT-CTP/5480",
    doi = "10.22323/1.430.0292",
    journal = "PoS",
    volume = "LATTICE2022",
    pages = "292",
    year = "2023"
}

@article{Assi:2024pdn,
    author = "Assi, Beno{\^\i}t and Lamm, Henry",
    title = "{Digitization and subduction of SU(N) gauge theories}",
    eprint = "2405.12204",
    archivePrefix = "arXiv",
    primaryClass = "hep-lat",
    reportNumber = "FERMILAB-PUB-24-0177-T",
    doi = "10.1103/PhysRevD.110.074511",
    journal = "Phys. Rev. D",
    volume = "110",
    number = "7",
    pages = "074511",
    year = "2024"
}

@article{Lamm:2024jnl,
    author = "Lamm, Henry and Li, Ying-Ying and Shu, Jing and Wang, Yi-Lin and Xu, Bin",
    title = "{Block encodings of discrete subgroups on a quantum computer}",
    eprint = "2405.12890",
    archivePrefix = "arXiv",
    primaryClass = "hep-lat",
    reportNumber = "USTC-ICTS/PCFT-24-15, FERMILAB-PUB-24-0242-T",
    doi = "10.1103/PhysRevD.110.054505",
    journal = "Phys. Rev. D",
    volume = "110",
    number = "5",
    pages = "054505",
    year = "2024"
}

@article{Harlow:2018tng,
    author = "Harlow, Daniel and Ooguri, Hirosi",
    title = "{Symmetries in quantum field theory and quantum gravity}",
    eprint = "1810.05338",
    archivePrefix = "arXiv",
    primaryClass = "hep-th",
    doi = "10.1007/s00220-021-04040-y",
    journal = "Commun. Math. Phys.",
    volume = "383",
    number = "3",
    pages = "1669--1804",
    year = "2021"
}

@article{Creutz:1979zg,
	Author = {Creutz, Michael and Jacobs, Laurence and Rebbi, Claudio},
	Doi = {10.1103/PhysRevD.20.1915},
	Journal = {Phys. Rev.},
	Pages = {1915},
	Reportnumber = {BNL-26307},
	Slaccitation = {%%CITATION = PHRVA,D20,1915;%%},
	Title = {{Monte Carlo Study of Abelian Lattice Gauge Theories}},
	Volume = {D20},
	Year = {1979}}

@article{Zohar:2014qma,
	Archiveprefix = {arXiv},
	Author = {Zohar, Erez and Burrello, Michele},
	Doi = {10.1103/PhysRevD.91.054506},
	Eprint = {1409.3085},
	Journal = {Phys. Rev.},
	Number = {5},
	Pages = {054506},
	Primaryclass = {quant-ph},
	Slaccitation = {%%CITATION = ARXIV:1409.3085;%%},
	Title = {{Formulation of lattice gauge theories for quantum simulations}},
	Volume = {D91},
	Year = {2015}}

@article{Bhanot:1981xp,
	Author = {Bhanot, G. and Rebbi, C.},
	Doi = {10.1103/PhysRevD.24.3319},
	Journal = {Phys. Rev.},
	Pages = {3319},
	Reportnumber = {Print-81-0392 (BNL)},
	Slaccitation = {%%CITATION = PHRVA,D24,3319;%%},
	Title = {{Monte Carlo Simulations of Lattice Models With Finite Subgroups of SU(3) as Gauge Groups}},
	Volume = {D24},
	Year = {1981}}

@article{Halimeh:2020xfd,
    author = "Halimeh, Jad C. and Ott, Robert and McCulloch, Ian P. and Yang, Bing and Hauke, Philipp",
    title = "{Robustness of gauge-invariant dynamics against defects in ultracold-atom gauge theories}",
    eprint = "2005.10249",
    archivePrefix = "arXiv",
    primaryClass = "cond-mat.quant-gas",
    doi = "10.1103/PhysRevResearch.2.033361",
    journal = "Phys. Rev. Res.",
    volume = "2",
    number = "3",
    pages = "033361",
    year = "2020"
}

@article{Weingarten:1981jy,
      author         = "Weingarten, Don",
      title          = "{Monte Carlo Evaluation of Hadron Masses in Lattice Gauge
                        Theories with Fermions}",
      journal        = "Phys. Lett.",
      volume         = "109B",
      year           = "1982",
      pages          = "57",
      doi            = "10.1016/0370-2693(82)90463-4",
      note           = "[,631(1981)]",
      reportNumber   = "IUHET-69",
      SLACcitation   = "%%CITATION = PHLTA,109B,57;%%"
}

@article{Weingarten:1980hx,
      author         = "Weingarten, D. H. and Petcher, D. N.",
      title          = "{Monte Carlo Integration for Lattice Gauge Theories with
                        Fermions}",
      journal        = "Phys. Lett.",
      volume         = "99B",
      year           = "1981",
      pages          = "333-338",
      doi            = "10.1016/0370-2693(81)90112-X",
      reportNumber   = "IUHET-61",
      SLACcitation   = "%%CITATION = PHLTA,99B,333;%%"
}

@article{Fradkin:1978dv,
    author = "Fradkin, Eduardo H. and Shenker, Stephen H.",
    doi = "10.1103/PhysRevD.19.3682",
    journal = "Phys. Rev. D",
    pages = "3682--3697",
    reportNumber = "SLAC-PUB-2238",
    title = "{Phase Diagrams of Lattice Gauge Theories with Higgs Fields}",
    volume = "19",
    year = "1979"
}

@article{Petcher:1980cq,
	Author = {Petcher, D. and Weingarten, D. H.},
	Doi = {10.1103/PhysRevD.22.2465},
	Journal = {Phys. Rev.},
	Pages = {2465},
	Reportnumber = {Print-80-0404 (INDIANA), IUHET-53},
	Slaccitation = {%%CITATION = PHRVA,D22,2465;%%},
	Title = {{Monte Carlo Calculations and a Model of the Phase Structure for Gauge Theories on Discrete Subgroups of SU(2)}},
	Volume = {D22},
	Year = {1980}}

@article{Ale:2024uxf,
    author = "Ale, Victor and Bauer, Nora M. and Jha, Raghav G. and Ringer, Felix and Siopsis, George",
    title = "{Quantum computation of SU(2) lattice gauge theory with continuous variables}",
    eprint = "2410.14580",
    archivePrefix = "arXiv",
    primaryClass = "hep-lat",
    reportNumber = "JLAB-THY-24-4217",
    doi = "10.1007/JHEP06(2025)084",
    journal = "JHEP",
    volume = "06",
    number = "084",
    year = "2025"
}

@article{Mueller:2024mmk,
    author = "Mueller, Niklas and Wang, Tianyi and Katz, Or and Davoudi, Zohreh and Cetina, Marko",
    title = "{Quantum computing universal thermalization dynamics in a (2{\,}+{\,}1)D Lattice Gauge Theory}",
    eprint = "2408.00069",
    archivePrefix = "arXiv",
    primaryClass = "quant-ph",
    reportNumber = "IQuS@UW-21-085, UMD-PP-024-08",
    doi = "10.1038/s41467-025-60177-7",
    journal = "Nature Commun.",
    volume = "16",
    number = "1",
    pages = "5492",
    year = "2025"
}

@article{Pardo:2024edu,
    author = "Pardo, Guy and Bender, Julian and Katz, Nadav and Zohar, Erez",
    title = "{Truncation-free quantum simulation of pure-gauge compact QED using Josephson arrays}",
    eprint = "2410.11413",
    archivePrefix = "arXiv",
    primaryClass = "hep-lat",
    reportNumber = "MIT-CTP/5817",
    doi = "10.1088/2058-9565/adce2a",
    journal = "Quantum Sci. Technol.",
    volume = "10",
    number = "3",
    pages = "035011",
    year = "2025"
}

@article{Raychowdhury:2018osk,
    author = "Raychowdhury, Indrakshi and Stryker, Jesse R.",
    title = "{Solving Gauss's Law on Digital Quantum Computers with Loop-String-Hadron Digitization}",
    eprint = "1812.07554",
    archivePrefix = "arXiv",
    primaryClass = "hep-lat",
    reportNumber = "INT-PUB-18-062",
    doi = "10.1103/PhysRevResearch.2.033039",
    journal = "Phys. Rev. Res.",
    volume = "2",
    number = "3",
    pages = "033039",
    year = "2020"
}

@article{Yamamoto:2020eqi,
    author = "Yamamoto, Arata",
    title = "{Real-time simulation of (2+1)-dimensional lattice gauge theory on qubits}",
    eprint = "2008.11395",
    archivePrefix = "arXiv",
    primaryClass = "hep-lat",
    doi = "10.1093/ptep/ptaa171",
    journal = "PTEP",
    volume = "2021",
    number = "1",
    pages = "013B06",
    year = "2021"
}

@article{Mathis:2020fuo,
    author = "Mathis, Simon V. and Mazzola, Guglielmo and Tavernelli, Ivano",
    title = "{Toward scalable simulations of Lattice Gauge Theories on quantum computers}",
    eprint = "2005.10271",
    archivePrefix = "arXiv",
    primaryClass = "quant-ph",
    doi = "10.1103/PhysRevD.102.094501",
    journal = "Phys. Rev. D",
    volume = "102",
    number = "9",
    pages = "094501",
    year = "2020"
}

@misc{Carena:2021ltu,
    author = "Carena, Marcela and Lamm, Henry and Li, Ying-Ying and Liu, Wanqiang",
    title = "{Lattice Renormalization of Quantum Simulations}",
    eprint = "2107.01166",
    archivePrefix = "arXiv",
    primaryClass = "hep-lat",
    reportNumber = "FERMILAB-PUB-21-222-T",
    month = "7",
    year = "2021"
}

@misc{Singh:2025sye,
    author = "Singh, Hersh",
    title = "{Ginsparg-Wilson Hamiltonians with Improved Chiral Symmetry}",
    eprint = "2505.20419",
    archivePrefix = "arXiv",
    primaryClass = "hep-lat",
    reportNumber = "FERMILAB-PUB-25-0330-T",
    month = "5",
    year = "2025"
}

@InProceedings{1998quant.ph..7064P,
author="P{\"u}schel, Markus
and R{\"o}tteler, Martin
and Beth, Thomas",
editor="Fossorier, Marc
and Imai, Hideki
and Lin, Shu
and Poli, Alain",
title="Fast Quantum Fourier Transforms for a Class of Non-abelian Groups",
booktitle="Applied Algebra, Algebraic Algorithms and Error-Correcting Codes",
year="1999",
publisher="Springer Berlin Heidelberg",
address="Berlin, Heidelberg",
pages="148--159",
isbn="978-3-540-46796-0"
}

@misc{Harmalkar:2020mpd,
    author = "Harmalkar, Siddhartha and Lamm, Henry and Lawrence, Scott",
    archivePrefix = "arXiv",
    collaboration = "NuQS",
    eprint = "2001.11490",
    month = "1",
    primaryClass = "hep-lat",
    reportNumber = "FERMILAB-PUB-20-045-T",
    title = "{Quantum Simulation of Field Theories Without State Preparation}",
    year = "2020"
}

@article{Hackett:2018cel,
    author = "Hackett, Daniel C. and Howe, Kiel and Hughes, Ciaran and Jay, William and Neil, Ethan T. and Simone, James N.",
    archivePrefix = "arXiv",
    doi = "10.1103/PhysRevA.99.062341",
    eprint = "1811.03629",
    journal = "Phys.\ Rev.\ A",
    number = "6",
    pages = "062341",
    primaryClass = "quant-ph",
    reportNumber = "FERMILAB-PUB-18-615-T",
    title = "{Digitizing Gauge Fields: Lattice Monte Carlo Results for Future Quantum Computers}",
    volume = "99",
    year = "2019"
}

@article{Bauer:2021gek,
    author = "Bauer, Christian W. and Grabowska, Dorota M.",
    title = "{Efficient representation for simulating U(1) gauge theories on digital quantum computers at all values of the coupling}",
    eprint = "2111.08015",
    archivePrefix = "arXiv",
    primaryClass = "hep-ph",
    reportNumber = "CERN-TH-2021-188",
    doi = "10.1103/PhysRevD.107.L031503",
    journal = "Phys. Rev. D",
    volume = "107",
    number = "3",
    pages = "L031503",
    year = "2023"
}

@article{Murairi:2024xpc,
    author = "Murairi, Edison M. and Sohaib Alam, M. and Lamm, Henry and Hadfield, Stuart and Gustafson, Erik",
    title = "{Highly-efficient quantum Fourier transformations for certain non-Abelian groups}",
    eprint = "2408.00075",
    archivePrefix = "arXiv",
    primaryClass = "quant-ph",
    reportNumber = "FERMILAB-PUB-24-0241-SQMS-T",
    doi = "10.1103/PhysRevD.110.074501",
    journal = "Phys. Rev. D",
    volume = "110",
    number = "7",
    pages = "074501",
    year = "2024"
}

@article{Gustafson:2024kym,
    author = "Gustafson, Erik J. and Ji, Yao and Lamm, Henry and Murairi, Edison M. and Perez, Sebastian Osorio and Zhu, Shuchen",
    title = "{Primitive quantum gates for an SU(3) discrete subgroup: \ensuremath{\Sigma}(36\texttimes{}3)}",
    eprint = "2405.05973",
    archivePrefix = "arXiv",
    primaryClass = "hep-lat",
    reportNumber = "FERMILAB-PUB-24-0132-SQMS-T, TUM-HEP-1506/24, FERMILAB-PUB-24-0132-SQMS-T; TUM-HEP-1506/24",
    doi = "10.1103/PhysRevD.110.034515",
    journal = "Phys. Rev. D",
    volume = "110",
    number = "3",
    pages = "034515",
    year = "2024"
}

@misc{Murairi:2022zdg,
    author = "Murairi, Edison M. and Cervia, Michael J. and Kumar, Hersh and Bedaque, Paulo F. and Alexandru, Andrei",
    title = "{How many quantum gates do gauge theories require?}",
    eprint = "2208.11789",
    archivePrefix = "arXiv",
    primaryClass = "hep-lat",
    month = "8",
    year = "2022"
}

@article{Hartung:2022hoz,
    author = "Hartung, Tobias and Jakobs, Timo and Jansen, Karl and Ostmeyer, Johann and Urbach, Carsten",
    title = "{Digitising SU(2) gauge fields and the freezing transition}",
    eprint = "2201.09625",
    archivePrefix = "arXiv",
    primaryClass = "hep-lat",
    doi = "10.1140/epjc/s10052-022-10192-5",
    journal = "Eur. Phys. J. C",
    volume = "82",
    number = "3",
    pages = "237",
    year = "2022"
}

@article{Farrell:2023fgd,
    author = "Farrell, Roland C. and Illa, Marc and Ciavarella, Anthony N. and Savage, Martin J.",
    title = "{Scalable Circuits for Preparing Ground States on Digital Quantum Computers: The Schwinger Model Vacuum on 100 Qubits}",
    eprint = "2308.04481",
    archivePrefix = "arXiv",
    primaryClass = "quant-ph",
    reportNumber = "IQuS@UW-21-060, NT@UW-23-13",
    doi = "10.1103/PRXQuantum.5.020315",
    journal = "PRX Quantum",
    volume = "5",
    number = "2",
    pages = "020315",
    year = "2024"
}

@article{Alexandru:2021jpm,
    author = "Alexandru, Andrei and Bedaque, Paulo F. and Brett, Ruair{\'\i} and Lamm, Henry",
    title = "{Spectrum of digitized QCD: Glueballs in a S(1080) gauge theory}",
    eprint = "2112.08482",
    archivePrefix = "arXiv",
    primaryClass = "hep-lat",
    reportNumber = "FERMILAB-PUB-21-683-T",
    doi = "10.1103/PhysRevD.105.114508",
    journal = "Phys. Rev. D",
    volume = "105",
    number = "11",
    pages = "114508",
    year = "2022"
}

@article{Horn:1979fy,
    author = "Horn, D. and Weinstein, M. and Yankielowicz, S.",
    title = "{Hamiltonian Approach to Z(N) Lattice Gauge Theories}",
    reportNumber = "TAUP-723-79",
    doi = "10.1103/PhysRevD.19.3715",
    journal = "Phys. Rev. D",
    volume = "19",
    pages = "3715",
    year = "1979"
}

@article{Ji:2020kjk,
    author = "Ji, Yao and Lamm, Henry and Zhu, Shuchen",
    collaboration = "NuQS",
    title = "{Gluon Field Digitization via Group Space Decimation for Quantum Computers}",
    eprint = "2005.14221",
    archivePrefix = "arXiv",
    primaryClass = "hep-lat",
    reportNumber = "FERMILAB-PUB-20-198-T",
    doi = "10.1103/PhysRevD.102.114513",
    journal = "Phys. Rev. D",
    volume = "102",
    number = "11",
    pages = "114513",
    year = "2020"
}

@article{Fromm:2022vaj,
    author = "Fromm, Michael and Philipsen, Owe and Winterowd, Christopher",
    title = "{Dihedral lattice gauge theories on a quantum annealer}",
    eprint = "2206.14679",
    archivePrefix = "arXiv",
    primaryClass = "hep-lat",
    doi = "10.1140/epjqt/s40507-023-00188-9",
    journal = "EPJ Quant. Technol.",
    volume = "10",
    number = "1",
    pages = "31",
    year = "2023"
}

@misc{osborne2024quantum,
    author = "Budde, Thea and Krsti{\'c} Marinkovi{\'c}, Marina and Barros, Joao C. Pinto",
    title = "{Quantum many-body scars for arbitrary integer spin in 2+1D Abelian gauge theories}",
    eprint = "2403.08892",
    archivePrefix = "arXiv",
    primaryClass = "hep-lat",
    doi = "10.1103/PhysRevD.110.094506",
    journal = "Phys. Rev. D",
    volume = "110",
    number = "9",
    pages = "094506",
    year = "2024"
}

@article{Carena:2024dzu,
    author = "Carena, Marcela and Lamm, Henry and Li, Ying-Ying and Liu, Wanqiang",
    title = "{Quantum error thresholds for gauge-redundant digitizations of lattice field theories}",
    eprint = "2402.16780",
    archivePrefix = "arXiv",
    primaryClass = "hep-lat",
    reportNumber = "USTC-ICTS/PCFT-24-06, FERMILAB-PUB-23-570-T",
    doi = "10.1103/PhysRevD.110.054516",
    journal = "Phys. Rev. D",
    volume = "110",
    number = "5",
    pages = "054516",
    year = "2024"
}

@article{Mathew:2022nep,
    author = "Mathew, Emil and Raychowdhury, Indrakshi",
    title = "{Protecting local and global symmetries in simulating (1+1)D non-Abelian gauge theories}",
    eprint = "2206.07444",
    archivePrefix = "arXiv",
    primaryClass = "hep-lat",
    doi = "10.1103/PhysRevD.106.054510",
    journal = "Phys. Rev. D",
    volume = "106",
    number = "5",
    pages = "054510",
    year = "2022"
}

@article{Irmejs:2022gwv,
    author = "Irmejs, Reinis and Banuls, Mari Carmen and Cirac, J. Ignacio",
    title = "{Quantum simulation of Z2 lattice gauge theory with minimal resources}",
    doi = "10.1103/PhysRevD.108.074503",
    journal = "Phys. Rev. D",
    volume = "108",
    number = "7",
    pages = "074503",
    year = "2023"
}

@article{Popov:2023xft,
    author = "Popov, Pavel P. and Meth, Michael and Lewenstein, Maciej and Hauke, Philipp and Ringbauer, Martin and Zohar, Erez and Kasper, Valentin",
    title = "{Variational quantum simulation of U(1) lattice gauge theories with qudit systems}",
    doi = "10.1103/PhysRevResearch.6.013202",
    journal = "Phys. Rev. Res.",
    volume = "6",
    number = "1",
    pages = "013202",
    year = "2024"
}

@misc{Ciavarella:2024fzw,
    author = "Ciavarella, Anthony N. and Bauer, Christian W.",
    title = "{Quantum Simulation of SU(3) Lattice Yang Mills Theory at Leading Order in Large N}",
    eprint = "2402.10265",
    archivePrefix = "arXiv",
    primaryClass = "hep-ph",
    month = "2",
    year = "2024"
}

@article{Kavaki:2024ijd,
    author = "Kavaki, Ali H. Z. and Lewis, Randy",
    title = "{From square plaquettes to triamond lattices for SU(2) gauge theory}",
    eprint = "2401.14570",
    archivePrefix = "arXiv",
    primaryClass = "hep-lat",
    doi = "10.1038/s42005-024-01697-4",
    journal = "Commun. Phys.",
    volume = "7",
    number = "1",
    pages = "208",
    year = "2024"
}

@misc{budde2024quantum,
    author = "Budde, Thea and Krsti{\'c} Marinkovi{\'c}, Marina and Barros, Joao C. Pinto",
    title = "{Quantum many-body scars for arbitrary integer spin in 2+1D Abelian gauge theories}",
    eprint = "2403.08892",
    archivePrefix = "arXiv",
    primaryClass = "hep-lat",
    doi = "10.1103/PhysRevD.110.094506",
    journal = "Phys. Rev. D",
    volume = "110",
    number = "9",
    pages = "094506",
    year = "2024"
}

@article{Alexandru:2020wrj,
    author = "Alexandru, Andrei and Basar, Gokce and Bedaque, Paulo F. and Warrington, Neill C.",
    title = "{Complex paths around the sign problem}",
    eprint = "2007.05436",
    archivePrefix = "arXiv",
    primaryClass = "hep-lat",
    doi = "10.1103/RevModPhys.94.015006",
    journal = "Rev. Mod. Phys.",
    volume = "94",
    number = "1",
    pages = "015006",
    year = "2022"
}

@inproceedings{Gustafson:2021jtq,
    author = "Gustafson, Erik and others",
    title = "{Large scale multi-node simulations of $\mathbb{Z}_2$ gauge theory quantum circuits using Google Cloud Platform}",
    booktitle = "{IEEE/ACM Second International Workshop on Quantum Computing Software}",
    eprint = "2110.07482",
    archivePrefix = "arXiv",
    primaryClass = "quant-ph",
    reportNumber = "FERMILAB-CONF-21-494-QIS",
    doi = "10.1109/QCS54837.2021.00012",
    month = "10",
    year = "2021"
}

@article{Alam:2021uuq,
    author = "Alam, M. Sohaib and Hadfield, Stuart and Lamm, Henry and Li, Andy C. Y.",
    collaboration = "SQMS",
    title = "{Primitive quantum gates for dihedral gauge theories}",
    eprint = "2108.13305",
    archivePrefix = "arXiv",
    primaryClass = "quant-ph",
    reportNumber = "FERMILAB-PUB-21-364-QIS-T",
    doi = "10.1103/PhysRevD.105.114501",
    journal = "Phys. Rev. D",
    volume = "105",
    number = "11",
    pages = "114501",
    year = "2022"
}

@misc{Grabowska:2022uos,
    author = "Grabowska, Dorota M. and Kane, Christopher and Nachman, Benjamin and Bauer, Christian W.",
    title = "{Overcoming exponential scaling with system size in Trotter-Suzuki implementations of constrained Hamiltonians: 2+1 U(1) lattice gauge theories}",
    eprint = "2208.03333",
    archivePrefix = "arXiv",
    primaryClass = "quant-ph",
    reportNumber = "CERN-TH-2022-133",
    month = "8",
    year = "2022"
}

@phdthesis{romers2007discrete,
  title={Discrete gauge theories in two spatial dimensions},
  author={Romers, JC},
  year={2007},
  school={Master’s thesis, Universiteit van Amsterdam}
}

@article{Kogut:1980qb,
    author = "Kogut, John B.",
    title = "{1/n Expansions and the Phase Diagram of Discrete Lattice Gauge Theories With Matter Fields}",
    reportNumber = "ILL-TH-80-01",
    doi = "10.1103/PhysRevD.21.2316",
    journal = "Phys. Rev. D",
    volume = "21",
    pages = "2316",
    year = "1980"
}

@misc{rajput2021quantum,
    author = "Rajput, Abhishek and Roggero, Alessandro and Wiebe, Nathan",
    title = "{Quantum error correction with gauge symmetries}",
    eprint = "2112.05186",
    archivePrefix = "arXiv",
    primaryClass = "quant-ph",
    reportNumber = "IQuS@UW-21-014",
    doi = "10.1038/s41534-023-00706-8",
    journal = "npj Quantum Inf.",
    volume = "9",
    number = "1",
    pages = "41",
    year = "2023"
}

@article{Ji:2022qvr,
    author = "Ji, Yao and Lamm, Henry and Zhu, Shuchen",
    collaboration = "NuQS",
    title = "{Gluon digitization via character expansion for quantum computers}",
    eprint = "2203.02330",
    archivePrefix = "arXiv",
    primaryClass = "hep-lat",
    reportNumber = "FERMILAB-PUB-21-496-T",
    doi = "10.1103/PhysRevD.107.114503",
    journal = "Phys. Rev. D",
    volume = "107",
    number = "11",
    pages = "114503",
    year = "2023"
}

@article{Gonzalez-Cuadra:2022hxt,
    author = "Gonz{\'a}lez-Cuadra, Daniel and Zache, Torsten V. and Carrasco, Jose and Kraus, Barbara and Zoller, Peter",
    title = "{Hardware Efficient Quantum Simulation of Non-Abelian Gauge Theories with Qudits on Rydberg Platforms}",
    eprint = "2203.15541",
    archivePrefix = "arXiv",
    primaryClass = "quant-ph",
    doi = "10.1103/PhysRevLett.129.160501",
    journal = "Phys. Rev. Lett.",
    volume = "129",
    number = "16",
    pages = "160501",
    year = "2022"
}

@article{Carena:2022kpg,
    author = "Carena, Marcela and Lamm, Henry and Li, Ying-Ying and Liu, Wanqiang",
    title = "{Improved Hamiltonians for Quantum Simulations of Gauge Theories}",
    eprint = "2203.02823",
    archivePrefix = "arXiv",
    primaryClass = "hep-lat",
    reportNumber = "FERMILAB-PUB-21-674-T",
    doi = "10.1103/PhysRevLett.129.051601",
    journal = "Phys. Rev. Lett.",
    volume = "129",
    number = "5",
    pages = "051601",
    year = "2022"
}

@article{PAIGE1994303,
title = {History and generality of the CS decomposition},
journal = {Linear Algebra and its Applications},
volume = {208-209},
pages = {303-326},
year = {1994},
issn = {0024-3795},
doi = {https://doi.org/10.1016/0024-3795(94)90446-4},
url = {https://www.sciencedirect.com/science/article/pii/0024379594904464},
author = {C.C. Paige and M. Wei}
}

@article{CHEN2013853,
title = {Qcompiler: Quantum compilation with the CSD method},
journal = {Computer Physics Communications},
volume = {184},
number = {3},
pages = {853-865},
year = {2013},
issn = {0010-4655},
doi = {https://doi.org/10.1016/j.cpc.2012.10.019},
url = {https://www.sciencedirect.com/science/article/pii/S0010465512003621},
author = {Y.G. Chen and J.B. Wang}
}

@article{PhysRevA.92.062317,
  title = {Optimal synthesis of multivalued quantum circuits},
  author = {Di, Yao-Min and Wei, Hai-Rui},
  journal = {Phys. Rev. A},
  volume = {92},
  issue = {6},
  pages = {062317},
  numpages = {7},
  year = {2015},
  month = {Dec},
  publisher = {American Physical Society},
  doi = {10.1103/PhysRevA.92.062317},
  url = {https://link.aps.org/doi/10.1103/PhysRevA.92.062317}
}

@article{Duncan2020graphtheoretic,
  doi = {10.22331/q-2020-06-04-279},
  url = {https://doi.org/10.22331/q-2020-06-04-279},
  title = {Graph-theoretic {S}implification of {Q}uantum {C}ircuits with the {ZX}-calculus},
  author = {Duncan, Ross and Kissinger, Aleks and Perdrix, Simon and van de Wetering, John},
  journal = {{Quantum}},
  issn = {2521-327X},
  publisher = {{Verein zur F{\"{o}}rderung des Open Access Publizierens in den Quantenwissenschaften}},
  volume = {4},
  pages = {279},
  month = jun,
  year = {2020}
}

@inproceedings{kissinger2020Pyzx,
    author = {Kissinger, Aleks and van de Wetering, John},
    title = {{PyZX: Large Scale Automated Diagrammatic Reasoning}},
    year = {2020},
    booktitle = {{\rm Proceedings 16th International Conference on} Quantum Physics and Logic, {\rm Chapman University, Orange, CA, USA., 10-14 June 2019}},
    editor = {Coecke, Bob and Leifer, Matthew},
    series = {Electronic Proceedings in Theoretical Computer Science},
    volume = {318},
    pages = {229-241},
    publisher = {Open Publishing Association},
    doi = {10.4204/EPTCS.318.14}
}
